# Broadband enhanced chirality with tunable response in hybrid plasmonic helical metamaterials

Ufuk Kilic[1,#], Matthew Hilfiker[1], Alexander Ruder[1], Rene Feder[2], Eva Schubert[1], Mathias Schubert[1,3,4], and Christos Argyropoulos[1,*]

[1]Department of Electrical and Computer Engineering, University of Nebraska-Lincoln, Lincoln, NE 68588, USA

[2]Fraunhofer Institute for Microstructure of Materials and Systems (IMWS), D-06120, Halle (Saale), Germany

[3]Department of Physics, Chemistry, and Biology, Linkoping University, 58183 Linkoping, Sweden

[4]Leibniz Institute for Polymer Research, Dresden, D-01005, Germany

[#]ufuk.kilic@huskers.unl.edu, *christos.argyropoulos@unl.edu

## Abstract

Designing broadband enhanced chirality is of strong interest to the emerging fields of chiral chemistry and sensing, or to control the spin orbital momentum of photons in recently introduced nanophotonic chiral quantum and classical optical applications. However, chiral light-matter interactions have an extremely weak nature, are difficult to be controlled and enhanced, and cannot be made tunable or broadband. In addition, planar ultrathin nanophotonic structures to achieve strong, broadband, and tunable chirality at the technologically important visible to ultraviolet spectrum still remain elusive. Here, we tackle these important problems by experimentally demonstrating and theoretically verifying spectrally tunable, extremely large, and broadband chiroptical response by nanohelical metamaterials. The reported new designs of all-dielectric and dielectric-metallic (hybrid) plasmonic metamaterials permit the largest and broadest ever measured chiral Kuhn's dissymmetry factor achieved by a large-scale nanophotonic structure. In addition, the strong circular dichroism of the presented bottom-up fabricated optical metamaterials can be tuned by varying their dimensions and proportions between their dielectric and plasmonic helical subsections. The currently demonstrated ultrathin optical metamaterials are expected to provide a substantial boost to the developing field of chiroptics leading to significantly enhanced and broadband chiral light-matter interactions at the nanoscale.

**Introduction**

Chirality or handedness is one of the most intriguing inherent properties of an object that cannot be made superimposable on its mirror image by simple symmetry operations (i.e., rotation or translation). This "symmetry breaking" phenomenon leads to different properties between left and right enantiomeric molecules that play a fundamental role in biology, chemistry, physics, and pharmacology.[1]-[3] However, chiral light-matter interactions have an extremely weak nature mainly due to the large-scale mismatch between the wavelength of incident circular polarized (chiral) light and the large size of natural chiral crystals (such as quartz or benzyl) or small size of chiral molecules (such as DNA or proteins).[4]-[5] The chiro-optical response, defined by the circular dichroism (CD) or optical activity of an absorbing medium, can be accurately evaluated by calculating the Kuhn's dissymmetry factor, also known as g-factor ($g_K$), that enables quantitative comparison among different chiral structures.[6]-[9] This dimensionless quantity is directly related to the broadly used CD metric; but is more appropriate to quantify the chiral response of plasmonic structures due to their inherent optical absorption. The exact formula that gives the $g_K$ factor is given by Equation 2 in Methods and is accurately computed by using generalized spectroscopic ellipsometry measurements. This unitless factor permits quantitative comparison of different chiroptical materials and takes into account both reflected and transmitted LH and RH circular polarized waves.

Recently, several efforts have been devoted to boost the weak chiral light-matter interactions in different frequency ranges by using artificially engineered structures, such as optical metamaterials, molecular self-assembled chiral plasmonic nanoparticles, and thin films based on enantiomer molecules.[7]-[30] However, all of these approaches were limited to relatively low and narrowband experimentally obtained g-factors, with the maximum values obtained mainly in infrared (IR) frequencies and not the visible or ultraviolet (UV) spectrum. In addition, the majority of the highest reported narrowband optical activity values are possible by using chiral molecules or other structures embedded in aqueous solutions,[8] which make them incompatible to on-chip nanophotonic-based circuit applications. Moreover, most of the chiral metamaterial structures are built based on top-down, complicated, and costly fabrication methods, prone to fabrication imperfections, limiting their flexibility, reproducibility, and, as a result, their practical applications. Finally, several chiral structures presented in the literature are based on single chiral resonator elements,[9]-[16] leading to weak output optical signals, mainly generated from extinction coefficient or luminescent measurements; and their chiral response is not broadband and cannot be tuned along different frequencies, spanning the UV, visible, and IR range. Hence, planar ultrathin nanophotonic structures to achieve extremely strong, broadband, and tunable chiral light-matter interactions at visible and UV frequencies still remain elusive.

In this work, we tackle this problem by demonstrating broadband and high $g_K$ factors generated by large-scale ultrathin helical metamaterials and accurately measured by ellipsometry. The glancing-angle deposition (GLAD) technique is used to fabricate these unique structures, which is an emerging bottom-up fabrication approach that is free of masks or templates and permits free choice of materials incorporated into three-dimensional (3D) nanostructures.[31]-[32] Notably, the chiroptical properties of various 3D nanostructures such as crescent,[33] pillar,[34] gammadion,[35] sphere,[36] and helical[37]-[48] were investigated in the literature at different frequency ranges than the visible/UV or by using alternative more complicated fabrication techniques. Unlike all previous relevant works based on GLAD,[42]-[48][48] here we present a unique bottom-up technique to achieve subsequent deposition of different dielectric (silicon/Si) and metallic (silver/Ag) materials leading to hybrid plasmonic helical metamaterials with a multiple number of subsegments and helices (as shown in Figure 1a) that achieve tunable chirality. We fully characterize their broadband and strong chiroptical response by using transmission and reflection mode spectroscopic ellipsometry. This is a more accurate measurement approach compared to the widely used alternative and simpler CD spectra-polarimeter analysis that usually cannot differentiate artifacts to circular dichroism originating from linear dichroism or other birefringence anisotropic effects and cannot provide information about the reflected chiral components of light.[8],[49]

The chiroptical response of single nanoscatterers that are dispersed into a liquid environment have been investigated before and high g-factor values comparable to our work were reported in[9], [13]. However, these previous results were based on extinction measurements that were subsequently used to estimate the macroscopic absorption, which are different from our accurate and explicit absorptance measurements. Hence, to the best of our knowledge, the experimentally measured broadband $g_K$ values presented in our work are one of the highest and broadest compared to any previously reported g-factor values in the visible to UV spectra obtained by various relevant structures that can be used in on-chip nanophotonic circuit applications.[18]-[20] Our experiments are verified by extensive finite element method (FEM) simulations that provide physical insights to understand the presented strong chiral response of the fabricated metamaterials and visualize their nanoscale chiral near-field distributions. The proposed ultrathin GLAD fabricated optical metamaterials can be directly integrated within nanophotonic system platforms and used in the next generation of chiral opto-electronic devices.[20] In addition, they are expected to be transformative to the emerging field of chiral quantum optics.[50]

**Results**
**Handedness effect on chirality**
The schematic representation of the hybrid helical metamaterials fabrication process based on the GLAD technique is presented in Figure 1a. The high resolution scanning electron microscopy (SEM) image of the resulted RH two turn all dielectric Si helical metamaterial is shown in Figure 1b. It should be noted that GLAD can only control the 3D assembly to a certain degree, similar to any nanofabrication method, which depends on the growth conditions and materials involved in the growth process. The main fabrication problems with the GLAD process are the anisotropic broadening growth, also known as fanning or shadowing effect, column bifurcation in the growth procedure, and competition between adjacent columns.[32], [51]-[53] By using the generalized Mueller matrix spectroscopic ellipsometry technique that takes into account both transmitted and reflected waves, we successfully measure (using Equation 2 in Methods) the spectral evolution of the Kuhn's g-factor for the RH Si helical metamaterial. The resulted broadband strong chiroptical response with a maximum value around ≈3 eV is seen in Figure 1d, clearly proving that the presented helical metamaterials have a pronounced optical chirality. Likewise, the GLAD fabrication of the LH two turn all dielectric Si helical metamaterial is performed and its $g_K$ spectra and SEM image are shown in Figure 1d and e, respectively. As expected by the near-mirror symmetry of the RH and LH helical metamaterials depicted in the SEM images, their g-factor spectra are nearly symmetric to each other. The minor asymmetry is attributed to fabrication imperfections during the GLAD process.

Furthermore, we fabricate, again based on the GLAD method (Figure 1a), a new hybrid metamaterial by incorporating plasmonic (Ag) subsegments into the Si helices leading to the formation of an array of heterostructure nanohelices. The incorporation of Ag subsegments results in multiple metal-dielectric interfaces along each nanohelix. The interfaces cause the formation of localized plasmons which, subsequently, affect the chiral response. Figure 1c and f show SEM images of RH and LH two turn Si-Ag hybrid plasmonic helical metamaterials, respectively. The resulting $g_K$ spectra of the new hybrid Si-Ag helical metamaterials exhibit a more narrowband and red shifted, but still enhanced, chiral response (see Figure 1d) compared to the Si chiral metamaterial. These spectra modifications originate from the plasmonic Ag subsegments inclusion. Notable, the increase in ohmic losses and absorption due to the incorporation of plasmonic subsegments within each nanohelix only marginally reduces the maximum g-factor values. Again, the minor asymmetry between LH and RH hybrid metamaterial g-factor responses shown in Figure 1d is attributed to GLAD fabrication imperfections. However, the handedness effect in the obtained strong chirality of the fabricated LH and RH metamaterials is clearly demonstrated by Figure 1d.

**All-dielectric helical metamaterials**
Initially, we focus our study on all-dielectric Si helical metamaterials that have low losses and a broadband pronounced chiral response. The obtained ellipsometric measured dissymmetry factor $g_K$, and cross section SEM images of the RH versions of these metamaterials for different helical turns are shown in Figure 2a. The number of helical turns has a significant impact on the $g_K$ factor values. The coupling of the circular polarized radiation to the presented metamaterials results in a significant enantioselective non-racemic ability that leads to a large differential circular polarized absorbance within an ultrabroad spectral range from ≈1.5 eV to ≈3.5 eV, as is evident in Figure 2a. The $g_K$ factor amplitude rapidly increases and its extremum red shifts as the number of helical turns increase from single to three helical turns and then saturates and becomes more narrowband for additional turns. This shift is mainly induced by the gradual increase in the major radius of the helical structure due to the anisotropic broadening GLAD growth effect, also known as the fanning phenomenon,[53] which results in a gradual rise in the packing density of the nanohelices as the turn number increases. The experimentally measured g-factor values are one of the highest and broadest compared to any previously reported g-factors obtained by relevant chiral nanophotonic structures (not including individual scatterers[9]) at visible to UV frequencies.[18]-[20]

Extensive simulations are performed to understand the presented strong chiral response of the fabricated metamaterials and the color density plot of the theoretically computed $g_K$ spectra versus the number of turns is depicted in Figure 2b. More details about the simulations are provided in Methods and Supplementary. The spectral response becomes narrower by increasing the number of turns, accompanied by a substantial resonant frequency red shift. The simulations are in excellent quantitative agreement with the experimental results shown in Figure 2a. The main discrepancy between the obtained g-factors is in their computed amplitude values that are found to be larger in the simulations. This discrepancy is attributed to the higher packing density of the GLAD fabricated samples and other geometrical disorder and fabrication imperfection effects that are not taken into account in the simulations. The average major and minor radii values together with the helical pitch sizes used in the simulations are obtained by a statistical analysis based upon various SEM images (see Supplementary Figure S1 and S2).

**Hybrid plasmonic helical metamaterials**
Next, we investigate the subsequent change of source material in the GLAD process from Ag to Si or vice versa that enables the creation of new hybrid plasmonic helical heterostructure metamaterials with multiple turns. In the current study, only one Ag subsegment is incorporated at each helical turn, consisting of 10% Ag and the remaining 90% is Si. The measured Kuhn's dissymmetry factor $g_K$ spectra for different helical turns is shown in Figure 3a. In contrast to the all-dielectric helical metamaterials (Figure 2a), the plasmonic Si-Ag helical metamaterials have the ability to narrow the chiral spectral response and control the amplitude of the Kuhn's dissymmetry factor as a function of the turn number, inducing only a minor spectral shift in the extremum point. We also perform simulations to obtain the evolution of Kuhn's g-factor spectra as a function of turn number for the same hybrid structure and the results are shown in Figure 3b. Again, despite the amplitude discrepancy between the theoretical and experimental results, the existence of a single dip with a very small spectral shift as a function of turn number shows an excellent quantitative agreement between theory and experiment. A complete theoretical study to inspect the effects of various geometrical parameters (minor and major radii or axial pitch) on $g_K$ spectra is presented in Supplementary Figure S6. It is interesting that the inclusion of an ultrathin plasmonic helical subsegment has a pronounced effect on the chiral response of the presented metamaterials.

**Enhanced chiral near fields**

The use of full wave simulations can help to not only identify and understand the resonant chiral response of the metamaterials but also visualize their enhanced chiral near-field distributions in the nanoscale, a very important property for chiral light-matter interaction applications [50]. The difference in electric ($\Delta \mathbf{E} = |\mathbf{E_{LH}}| - |\mathbf{E_{RH}}|$) and magnetic ($\Delta \mathbf{B} = |\mathbf{B_{LH}}| - |\mathbf{B_{RH}}|$) field values induced by LH and RH circular polarized light illuminations are computed by using FEM-based simulations; and the results are shown in the right panels of Figure 2c and Figure 3c. The results are plotted at the spectral locations of the g-factor minima (red stars in Figure 2b and Figure 3b), where the chiral response of the metamaterials is the strongest. Those points correspond to the four turn all-dielectric Si (Figure 2b) and hybrid plasmonic Si-Ag (Figure 3b) helical metamaterials, respectively. These two points were selected in order to provide further theoretical physical insights to the experimental results presented in Figure 2a and Figure 3a. In the left side of Figure 2c and Figure 3c, the schematic illustrations of the unit cells used in the simulations and the SEM images of an isolated Si or Si-Ag helix are also depicted at the top and bottom sections, respectively. The color density distributions of $\Delta \mathbf{E}$ and $\Delta \mathbf{B}$ along a two-dimensional (2D) cut slice (with position shown as red hatched area) are demonstrated at the right top and bottom panels, respectively, in Figure 2c and Figure *3*c. The color scale displays the normalized magnitude of the field difference with red and blue colors indicating the maximum positive and negative differences, respectively. The inner hollow part of the Si helical metamaterial (Figure 2c) has pronounced $\Delta \mathbf{E}$ positive values (red color). However, the inner hollow part of the same helix has intense $\Delta \mathbf{B}$ negative values (blue color); whereas the outer helix sides have positive values (red color). Hence, intense asymmetric chiral near-field distributions induced by LH and RH circular polarized light illuminations are obtained in perfect agreement to the enhanced macroscopic chiral response demonstrated by the high g-factor values at the resonance of this metamaterial. These pronounced near-field enhancement distributions can be used to enhance the chiral light-matter interactions at the nanoscale leading to exciting new applications [50]. On the contrary, the 2D color density distributions of $\Delta\mathbf{E}$ and $\Delta\mathbf{B}$ for the hybrid Si-Ag helical metamaterial (Figure 3c) clearly demonstrate that the difference is more pronounced in the nanoscale vicinity of the plasmonic Ag subsegments, proving that the g-factor resonance dip effect is indeed due to localized plasmons that can control the enhanced chiral response.

**Tunable enhanced chirality**

To further demonstrate the spectrally tunable chiral response of the proposed hybrid helical metamaterials, i.e., their ability to engineer the spectral location of the g-factor resonances, we perform another systematic experimental study by changing the number of Ag plasmonic subsegments at each helical turn. More specifically, we fabricate based on GLAD hybrid helical metamaterials with fixed two turns, where each helical turn includes: (i) zero (only Si), (ii) one, (iii) two, and (iv) three Ag subsegments. Each Ag subsegment has an extremely nanoscale length of approximately 20 nm and similar diameter to the nanohelix. Next, we use the spectroscopic ellipsometry optical characterization to successfully measure their g-factor spectra as shown in Figure 4a together with their SEM images.

The spectral evolution of the Kuhn's dissymmetry factor versus different Ag subsegments at each Si helix is outlined in Figure 4a. It exhibits a single dip behavior with lower amplitude, where the extremum is spectrally blue shifted as the Si amount in the helical heterostructure is increased (less metal). In addition, the response of Figure 4a is more broadband and can be further tuned compared to the single Ag subsegment at each helix case shown in Figure 3a. We also perform a relevant systematic theoretical study involving FEM-based simulations with results shown in Figure 4b. As the ratio of Si in the helical heterostructure increases, the observed dip in the theoretical Kuhn's g-factor values starts to blue shift which is in excellent qualitative agreement with the experimental findings presented in Figure 4a. Our simulations can accurately compute the g-factor spectral position but fail at some cases to precisely predict the g-factor magnitude. This discrepancy is more pronounced for multi-segment elongated helical metamaterials and is a direct consequence of the GLAD fabrication imperfections, in particular the fanning/shadowing effect, bifurcation in the growth procedure, and competition between adjacent very closely packed helices.[32], [51]-[53] Hence, the structural disorder increases for larger segment number, which results to higher overall effective absorption from the experimentally obtained elongated helical metamaterials that, in turn, reduces the absorption-based measured chiral g-factor.

Increasing the number of subsegments per turn leads to an extreme confinement of light into the multiple metal-dielectric interfaces. This is demonstrated by the 3D color density distribution of the electric field enhancement shown in Figure 4c for an isolated hybrid Si-Ag helix composed of three Ag subsegments per turn monitored at the photon energy indicated by the red star in Figure 4b. The green highlighted area shown as inset in Figure 4c clearly demonstrates the highly confined and strong plasmonic field enhancement along the metal-dielectric interfaces. The chiral near fields are always localized and enhanced in the vicinity of the plasmonic Ag subsegments, a unique feature of the currently presented hybrid plasmonic helical metamaterials.

Finally, another more extensive theoretical parametric study based on simulations is performed by employing the hybrid helical metamaterials now with fixed four turns. Each helical turn includes a single Ag subsegment on top of the Si subsegment, but the rate of the Si segment ($f_{Si}=1-f_{Ag}$) varies from 0 (i.e., all-metallic plasmonic metamaterial made only of Ag helices) to 1 (i.e., all-dielectric nanostructure made only of Si helices). As seen in Figure 5, a strong resonance emerges in the computed Kuhn's g-factor ($g_K$) spectra that can be efficiently tuned by varying each helix Si ratio. Interestingly, Figure 5 also predicts that the g-factor amplitude will be maximum in the case of all-metallic (plasmonic) Ag helices but at lower frequencies compared to the currently presented hybrid metamaterials. The theoretical results in Figure 5 clearly present the unique tunable properties of the proposed hybrid plasmonic-dielectric helical metamaterials. The incorporation of plasmonic material in each helix leads to a substantial redshift and enhancement in the chiral response (g-factor) of the presented new hybrid metamaterials. Localized plasmon resonances along the metallic subsegments of each nanohelix cause multiple local extremum points in the computed g-factor spectrum. Mutual coupling between the neighboring helices, leading to collective plasmonic resonant modes, is an additional reason of the high g-factor values of Figure 5 for increased metallic segment ratio. The experimental realization of pure plasmonic nanohelical metamaterials will consist our future work, since metal growth makes the GLAD method even more prone to fabrication irregularities, such as fanning/shadowing effect, bifurcation in the growth procedure, and competition between adjacent very closely packed helices.[32], [51]-[53]

**Discussion**

The presented large-scale ultrathin helical metamaterials, fabricated by using the GLAD bottom-up technique and accurately measured by ellipsometry, can achieve a full control of chirality at the nanoscale in a broad frequency range including the technologically important visible to UV spectrum. The demonstrated broadband and tunable chiral response generated by the all-dielectric and hybrid plasmonic helical metamaterial designs will lead to new potential opportunities in the development of the next generation chiroptical nanophotonic devices.[20] Each subwavelength scale helix in the hybrid design is divided into multiple subsegments causing a unique chiral plasmonic response based on numerous metal-dielectric interfaces in the nanoscale. This new fabrication process results in localized plasmons that can cause a substantial spectral tuning of the enhanced chirality ranging from visible to UV frequencies. The currently presented work sets new benchmarks in the assembly of ultrathin plasmonic chiral metamaterials which are poised to efficiently control and enhance the chiral light-matter interactions at the nanoscale. It provides a comprehensive road map for designing hybridized plasmonic helical metamaterials with unprecedentedly high and broadband chiroptical properties that can be used in a plethora of diverse emerging applications, such as in the design of ultrathin polarization filters,[21] chiral sensors,[54]-[55] circular polarized single- or multi-photon radiation sources,[56]-[57] and directional spin-dependent nanophotonic waveguides.[58]-[59]

## Methods
### GLAD fabrication process
GLAD is an emerging bottom-up physical vapor deposition fabrication technique that is free of masks or templates and permits free choice of materials incorporated into 3D nanostructures. It is based on both high precision sample stage manipulation and oblique angle particle flux deposition mechanisms. It can lead to the fabrication of well-controlled 3D morphologies at the nanoscale due to competitive nucleation kinetics, geometric shadowing, and limitation of adatom surface diffusion processes. Thus, these properties make the GLAD technique highly advantageous compared to other nanofabrication methods due to its fast, simple, cost-effective, and scalable mass-production capability.

Here, we use a custom built ultra-high vacuum GLAD system with a base pressure of $1.0 \times 10^{-9}$ mbar, leading to either continuous Si or periodically alternating segments of Si and Ag helical nanostructures deposition. By placing the glass substrate at an extreme oblique angle of 85.5º to the source of incoming/incident vapor flux and by rotating the sample stage either clockwise or counter clockwise, we successfully create RH and LH helical metamaterials, as schematically illustrated in Figure 1a. An electron beam of V=8.8 kV and $I_{avr} \approx 180$ mA is used to evaporate Si, while for the Ag a beam of V=8.9 kV and $I_{avr} \approx 215$ mA is used. These parameters result in an average chamber pressure of $2.0 \times 10^{-7}$ mbar for Ag growth and $1.2 \times 10^{-7}$ mbar for Si growth. The current values vary during the deposition process in order to keep a constant deposition rate. A quartz crystal micro-balance (QCM) is integrated into the reaction chamber to monitor the material growth rate and total amount of deposited material during the helical structure fabrication process. Each subsegment helical length and, thereby, the total thickness of the helical heterostructure is controlled by the deposition time. The typical deposition rates are 0.1 Å/s for Si and 2.5 Å/s for Ag. The rotation speed of the sample stage is kept constant at 24 rpm for all hybrid helical metamaterials. An electronic shutter system is utilized to grow each periodic layer for the same time interval. The Ag growth is controlled by the total amount of Ag deposited on the QCM for a better control over the thickness of these extremely thin plasmonic layers. The structural parameters (total thickness, major and minor radii, pitch length) of the fabricated helical metamaterials are determined from high-resolution SEM images obtained by using the FEI Helios NanoLab 660 SEM instrument.

**Dissymmetry factor measurements**

Typically, the g-factor values reported in the literature for different type of thin films, nanomaterials, and nanostructures depend upon the sample thickness and orientation.[18]-[20] Commercial CD spectrographs are usually used to compute the g-factor but this simplistic measurement approach, which is suitable to measure simple isotropic samples, includes artifacts originating from linear dichroism or other anisotropic birefringence effects that are irrelevant to circular dichroism and cannot provide information about the chiral reflected waves.[8],[49] The currently used generalized spectroscopic ellipsometry[60] can correctly differentiate the linear from circular dichroism in both transmission and reflection by computing all of the so-called Mueller matrix elements[61] from which circular and linear dichroism properties can be accurately derived for any given sample under investigation. The Mueller matrix elements describe the effect of the sample on any incoming different polarized electromagnetic wave. This process is represented by the relation between the input and output Stokes vector parameters given by the formula:

$$\begin{pmatrix} S_0 \\ S_1 \\ S_2 \\ S_3 \end{pmatrix}_{out} = \begin{pmatrix} M_{11} & M_{12} & M_{13} & M_{14} \\ M_{21} & M_{22} & M_{23} & M_{24} \\ M_{31} & M_{32} & M_{33} & M_{34} \\ M_{41} & M_{42} & M_{43} & M_{44} \end{pmatrix} \begin{pmatrix} S_0 \\ S_1 \\ S_2 \\ S_3 \end{pmatrix}_{in}. \qquad (1)$$

The Stokes vector components are defined as: $S_0 = I_p + I_s = I_+ + I_- = I_{tot}$, $S_1 = I_p - I_s$, $S_2 = I_{45} - I_{-45}$, and $S_3 = I_+ - I_-$. Here, $I_{tot}, I_p, I_s, I_{45}, I_{-45}, I_+, \text{ and } I_-$ denote the total, p-, s-, +45,-45 , LH (+), and RH (-) linear and circular polarized light intensities, respectively [60].

The Kuhn's g-factor definition is derived by the absorbance of LH (+) and RH (-) circularly polarized light (A- and A+, respectively). It is calculated by the difference of these two absorbances, which is analogous to the circular dichroism metric, divided by their summation. It can be accurately computed by the Mueller matrix elements measured at normal incidence in both transmission (T) and reflection (R) configuration set-ups (i.e., two measurements are required). More details on how the various Mueller matrix elements are connected to the circular polarized absorbance are provided in Supplementary Section 2.1. The derived formula of the Kuhn's g-factor ($g_K$) is the following:

$$g_K = 2\frac{A_- - A_+}{A_- + A_+} = 2\frac{M_{14}^R + M_{14}^T}{1 - M_{11}^R - M_{11}^T}. \tag{2}$$

The Kuhn's g-factor definition given by Equation (2) is a complete and, as a result, correct and accurate metric to precisely characterize the chiral asymmetry in the absorbance rate of free-standing nanostructures that can transmit and reflect the incident light.

**Simulations**

Extensive FEM-based simulations are performed by using the RF module of COMSOL Multiphysics software. This module computes the frequency dependent full-wave electromagnetic solutions of Maxwell's equations. The incoming plane wave with a known circular polarization (either RH or LH) interacts with the helical metamaterials and both reflected and transmitted plane waves are computed by simulations. Moreover, the 3D and 2D electric and magnetic field distributions in and around the helical nanostructures are calculated and plotted. The S-parameters, also known as scattering parameters, are computed by using the multiple port definition.[32] The cross and co-polarized transmittance and reflectance coefficients are calculated from the obtained scattering parameters. The derived total reflectance and transmittance values contain both cross and co-polarized components. The spectral dependence of absorbance and circular dichroism ($g_K$) is calculated based on the computed total reflectance and transmittance values. All simulations are performed at normal angle of incidence, similar to the experimental procedure. A square unit cell is used with periodic boundary conditions incorporated to account for the interactions between the neighboring nanohelices. Note that the simulated helices are perfect (no roughness or disorder is considered) and are packaged in a less dense formation compared to the experimental nanostructures. This is the main reason why a perfect quantitative agreement between theoretical and experimental results is not obtained. An adaptive extremely fine mesh is employed with a maximum element size of 1.5 nm. The wavelength dependence of the Ag and Si subsegments dispersive dielectric constants are taken from experimental data.[62] FEM modeling is also performed to investigate the effect of Si crystallinity by using the dielectric functions of single crystalline[63] and polycrystalline[64] Si. We observe that the former gives rise to a larger g-factor maximum value ($g_{K;max} \approx 1.05$) compared to the latter dielectric function ($g_{K;max} \approx 0.4$). The resulting color density plot of $g_K$ spectra as a function of number of turns for the single crystalline Si is shown in Supplementary Figure S7a.

**Acknowledgements**

This work was partially supported by the National Science Foundation under award number DMR 1808715, Air Force Office of Scientific Research under award number FA9550-18-1-0360, Nebraska Materials Research Science and Engineering Center under award number DMR 1420645, Swedish Knut and Alice Wallenbergs Foundation supporting grant titled 'Wide-bandgap semi-conductors for next generation quantum components', and American Chemical Society/Petrol Research Fund. R. F. acknowledges the German Research Foundation (DFG) award FE 1532/1-1. C. A. acknowledges partial support by the Office of Naval Research Young Investigator Program (ONR YIP) under award number N00014-19-1-2384. M. S. acknowledges the University of Nebraska Foundation and the J. A. Woollam Foundation for financial support.

**Conflict of interest:**

The authors declare that they have no competing interests.

**Data and materials availability:**

All data needed to evaluate the conclusions in the paper are present in the paper and/or the Supplementary Materials.

# Figures

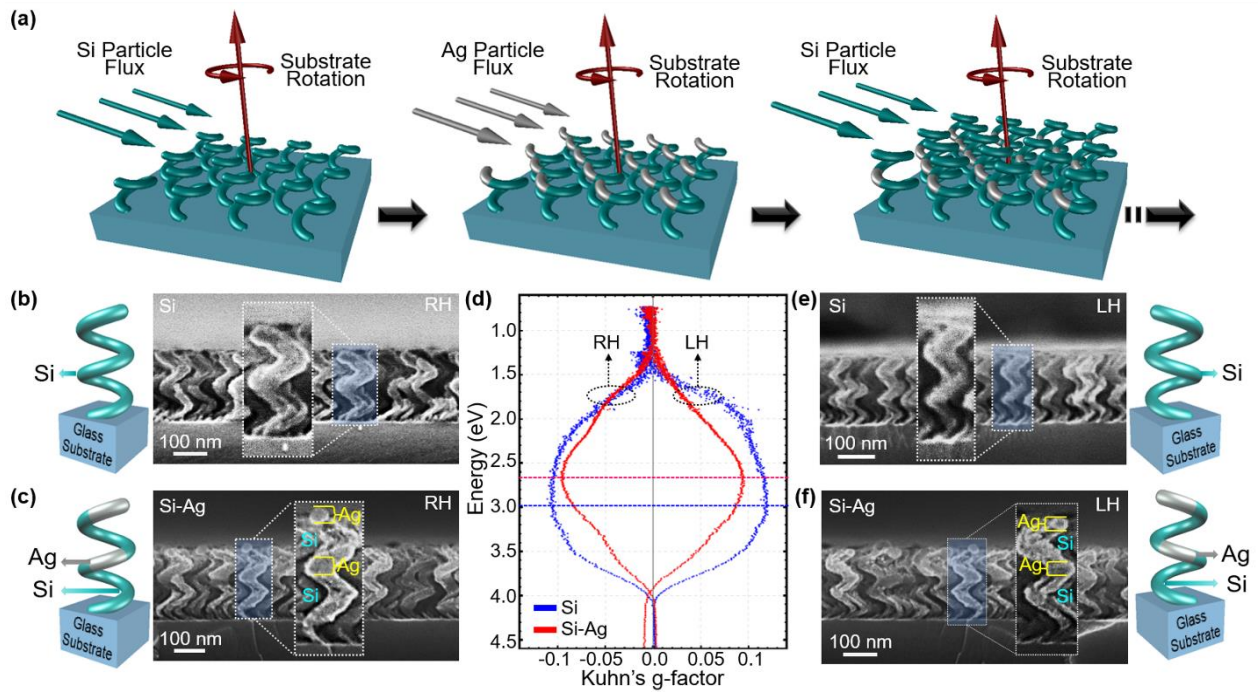

**Figure 1** (a) GLAD fabrication procedure of a chiral heterostructure nanohelices array. (b)-(f) Cross section SEM images together with their schematics of (b), (c) RH and (e), (f) LH dielectric Si (top: (b), (e)) and hybrid plasmonic Si-Ag (bottom: (c), (f)) helices, respectively. (d) Spectral evolution of experimentally obtained enhanced and broadband Kuhn's dissymmetry g-factors ($g_K$) for RH and LH two turn all-dielectric Si helical metamaterials (blue lines). The Kuhn's g-factors of RH and LH plasmonic two turn Si-Ag helical metamaterials are also presented (red lines) in the same plot to demonstrate the spectral tunability and narrower bandwidth due to the metallic subsegment inclusion.

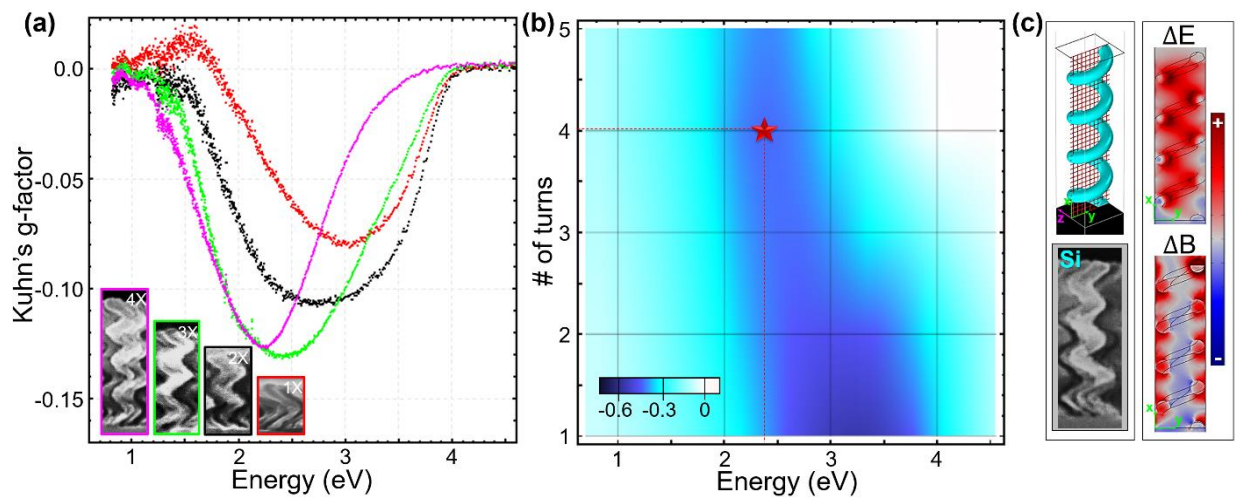

**Figure 2** (a) Experimentally measured and (b) theoretically computed spectral evolution of the Kuhn's dissymmetry factor for different helical turns in the case of RH all-dielectric Si helical metamaterials. (c) Left panels: schematic of an isolated Si helix (top) and its SEM image (bottom). Right panels: the asymmetric responses of the induced electric (top; ΔE) and magnetic (bottom; ΔB) circular polarized fields plotted on a 2D slice and monitored at the photon energy indicated by the red star in (b).

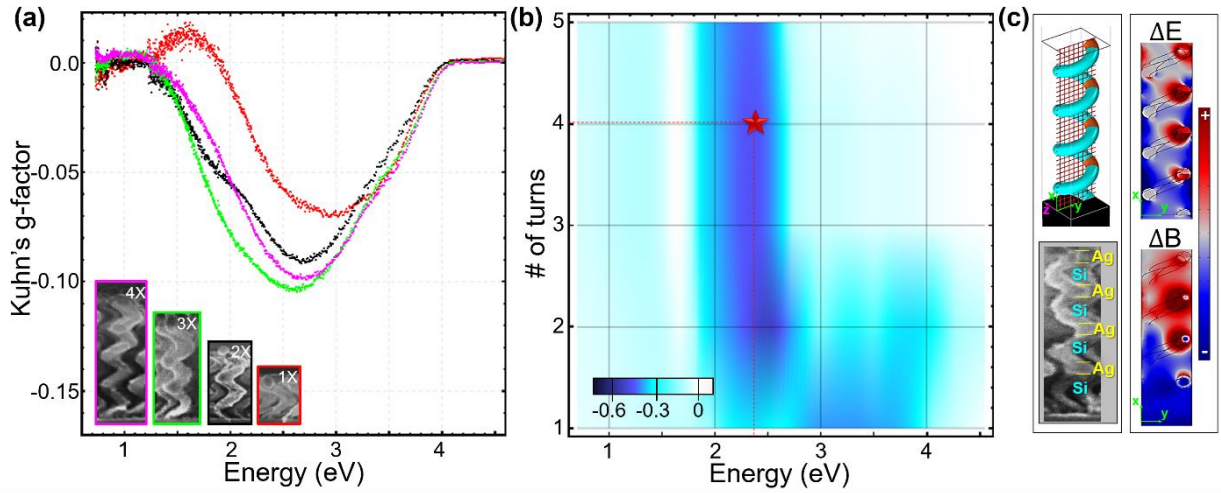

**Figure 3** (a) Experimentally measured and (b) theoretically computed spectral evolution of the Kuhn's dissymmetry factor for different helical turns in the case of RH hybrid plasmonic Si-Ag helical metamaterials. (c) Left panels: schematic of an isolated Si-Ag hybrid helix (top) and its SEM image (bottom). Right panels: the asymmetric responses of the induced electric (top; ΔE) and magnetic (bottom; ΔB) circular polarized fields plotted on a 2D slice and monitored at the photon energy indicated by the red star in (b).

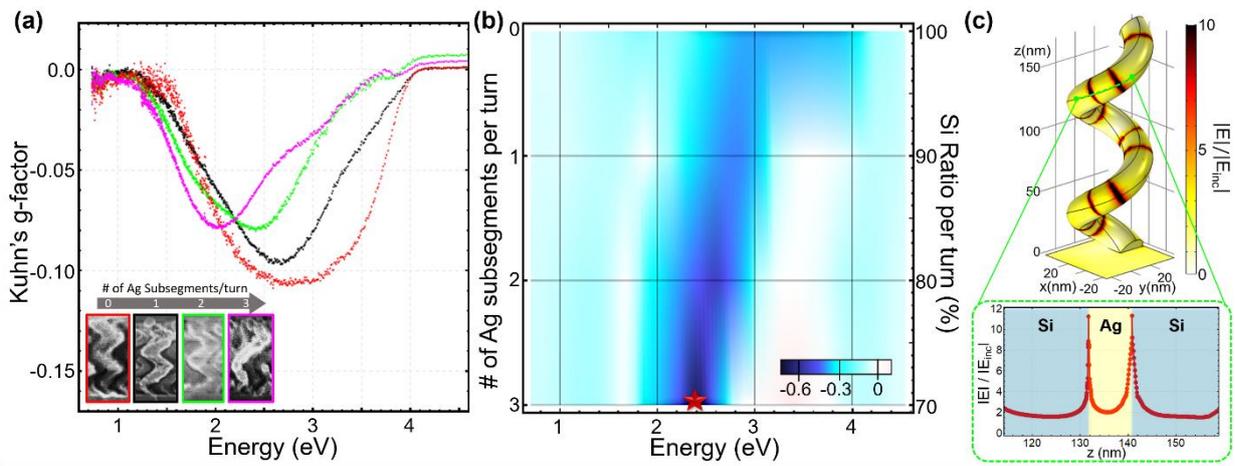

**Figure 4** (a) Experimentally measured and (b) theoretically computed spectral evolution of the Kuhn's dissymmetry factor for different silver subsegments per helix turn in the case of two turn RH hybrid plasmonic Si-Ag helical metamaterials. (c) 3D color density distribution of the electric field enhancement for an isolated hybrid Si-Ag helix composed of three Ag subsegments per turn monitored at the photon energy indicated by the red star in (b). The green highlighted area shown as inset in (c) clearly demonstrates the plasmonic field enhancement along the metal-dielectric interfaces.

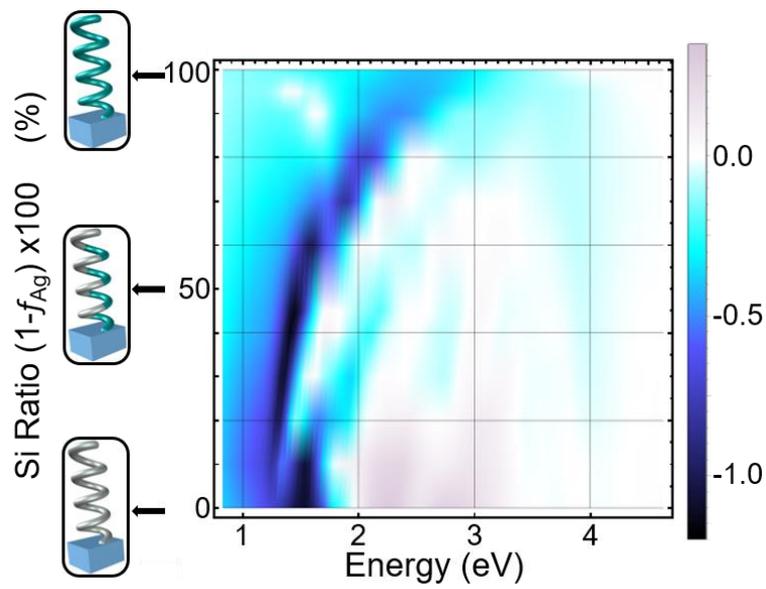

**Figure 5** Theoretically computed spectral evolution of the Kuhn's dissymmetry factor for different Si ratios per helix turn in the case of four turn RH Si-Ag hybrid helical metamaterials.

## Supporting Information
### 1. Fabrication process details

As stated in the main paper, we employed a custom built ultra-high vacuum glancing angle deposition (GLAD) instrument in order to bottom-up fabricate all-dielectric or plasmonic heterostructure helical metamaterials. In this section, we demonstrate a statistical study based on a scanning electron microscopy (SEM) high resolution image analysis with the goal to derive the various structural parameters of the helical nanostructures, such as minor and major radii and pitch height. These values serve as critical inputs to the performed finite element method (FEM) simulations. The results of this study are presented in Figure S1 and S2.

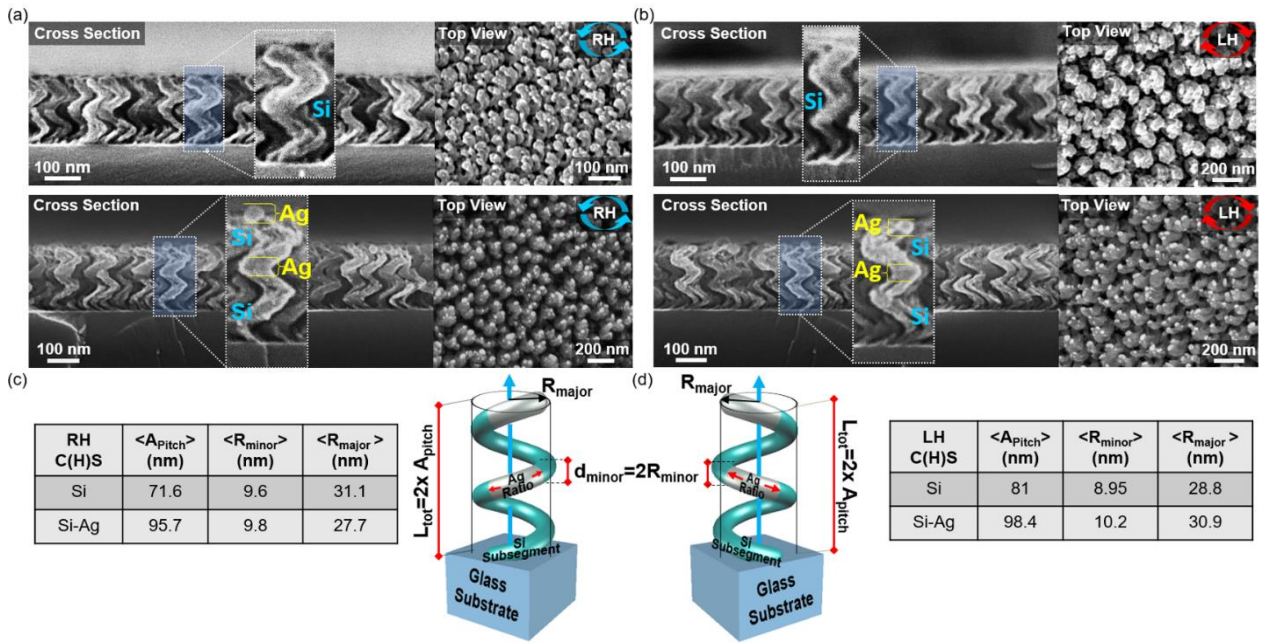

**Figure S1** Helical metamaterials fabricated by GLAD. (a)-(b) SEM cross-section and top view images of (a) RH and (b) LH two helical turn (top) Si and (bottom) Si-Ag metamaterials. (c)-(d) Schematic illustrations of the unit cell of (c) RH and (d) LH hybrid nanohelices and their main geometrical parameters (total length ($L_{tot}$), minor radius ($R_{minor}$), major radius ($R_{major}$), and pitch size $A_{pitch}$). The average values of these parameters are obtained from a statistical analysis of their geometrical dimensions taken by SEM images and are summarized into the two tables for each handedness case.

It is worth mentioning that nanocolumnar thin films fabricated by using the GLAD technique, similar to the current helical metamaterials, might exhibit changes in their structural parameters (especially column radius, slanting angles, etc.) during their fabrication mainly due to an anisotropic broadening effect, also known as the fanning phenomenon.[1] Figure S2a and S2b show the high-resolution cross section SEM images of RH all-dielectric Si and plasmonic Si-Ag helical metamaterials, respectively, as a function of number of turns. In order to investigate how the structural parameters of the fabricated helical metamaterials evolve as the number of turns increases from one to four, we performed a statistical study to extract the evolution of the structural parameters that is summarized in Figure S2c.

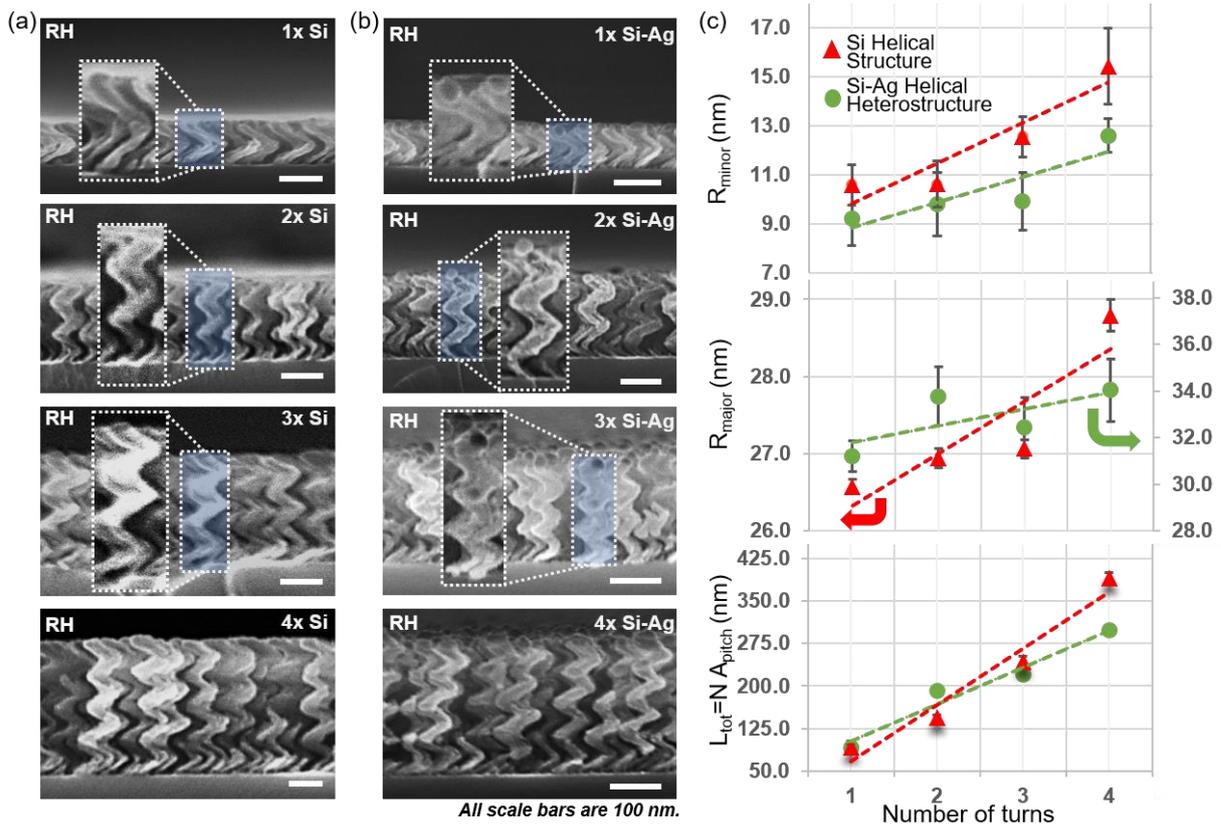

**Figure S2** Collection of high-resolution SEM images of (a) Si and (b) Si-Ag helical metamaterials with different turns. (c) Minor radius ($R_{minor}$), major radius ($R_{major}$), and total thickness ($L_{tot}$) as a function of number of helical turns obtained from the SEM images. While the vertical bars indicate the standard deviation intervals, the dashed green and red lines are the linear fits to the average measured values of $R_{minor}$, $R_{major}$, and $L_{tot}$.

## 2. Ellipsometric measurement analysis details

Here, we report the details of the used experimental technique that was developed to accurately measure the circular dichroism and therefore the Kuhn's dissymmetry factor shown in the main paper. The experimental set-up for both the reflection and transmission mode measurements are schematically illustrated in Figure S3. The spectral evolution of the 4x4 Mueller matrix elements in transmission and reflection modes for selected structures (all-dielectric Si and hybrid plasmonic Si-Ag helical metamaterials) are presented in Figure S4 and S5.

## 2.1. Circular polarized absorbance calculations

By using the Mueller matrix shown in Equation (1) in the main paper,[1],[3]-[5] one can obtain the Stokes elements parameters by the formula: $S_{i-1}^{out,\pm} = \sum_{j=1}^{4} M_{ij} S_{i-1}^{in,\pm}$, where i=1,2,3, and 4 [6]. The Stokes elements relevant to circular polarized radiation are $S_0$ and $S_3$. The other Stokes vector components ($S_1$ and $S_2$) are only related to linear polarized radiation and can be assumed to be zero when circular polarized waves are solely used. Hence, the transmitted ($I_{\pm}^T$) and reflected ($I_{\pm}^R$) circular polarized light intensities are derived to be:

$$I_{\pm}^T = (M_{11}^T \pm M_{14}^T) I_{+}^{in} \qquad I_{\pm}^R = (M_{11}^R \pm M_{14}^R) I_{+}^{in} \qquad \text{S1}$$

where $I_{\pm}^{in}$ is the input circular polarized light intensity. The circular polarized reflectance ($R_{\pm}$) and transmittance ($T_{\pm}$) can be obtained in terms of Mueller matrix elements as follows:

$$R_{\pm} = I_{\pm}^R / I_{\pm}^{in} = (M_{11}^R \pm M_{14}^R) \Rightarrow \begin{matrix} R_+ - R_- = 2 M_{14}^R \\ R_+ + R_- = 2 M_{11}^R \end{matrix}$$

$$T_{\pm} = I_{\pm}^T / I_{\pm}^{in} = (M_{11}^T \pm M_{14}^T) \Rightarrow \begin{matrix} T_+ - T_- = 2 M_{14}^T \\ T_+ + T_- = 2 M_{11}^T \end{matrix} \qquad \text{S2}$$

Energy conservation mandates the relationships between absorbance, transmittance, and reflectance for the LH and RH circular polarizations to be: $A_{\pm} = 1 - R_{\pm} - T_{\pm}$. It should be noted that the used total reflectance and transmittance include both co- and cross-polarization circular polarization components. The differential absorbance ($A_- - A_+$) and the total absorbance ($A_- + A_+$) can be obtained by using Equation (S2):

$$A_- - A_+ = (1 - R_- - T_-) - (1 - R_+ - T_+) = \underbrace{(T_+ - T_-)}_{2 M_{14}^T} + \underbrace{(R_+ - R_-)}_{2 M_{14}^R}$$

$$A_- + A_+ = (1 - R_- - T_-) + (1 - R_+ - T_+) = 2 - \underbrace{(T_+ + T_-)}_{2 M_{11}^T} - \underbrace{(R_+ + R_-)}_{2 M_{11}^R} \qquad \text{S3}$$

Equations (S3) are used in the Methods Section of the main paper to compute the g-factor which characterizes the chirality of the presented metamaterials.

## 2.2. Details of reflection mode analysis

While the transmission mode ellipsometric measurements are straightforward to be performed (see Figure S3a), the reflection mode analysis is more complicated. The measured reflected Mueller matrix data includes the results obtained by the experiments performed with the mirrors on wedge stage reflection mode set-up shown in Figure S3b. In order to accurately extract the reflected Mueller matrix elements of the sample, we successfully built an optical system model based on the optical model analysis software WVASE-32, which is commercially available to our spectroscopic ellipsometry instruments. During the near normal reflection mode spectroscopic ellipsometry data acquisition, a $SiO_2$ thin film is used as the reference sample. In the first step of data acquisition, the ellipsometer calibrated the light source intensity with respect to the reflection of light from the first gold (Au) Mirror-1 to the $SiO_2$ thin film sample and, then, to the second Au Mirror-2. In the second step, the $SiO_2$ thin film sample is replaced with the various nanohelical metamaterial samples and the Mueller matrix data acquisition is performed, accordingly. The mirrors on the wedge stage have angles $\varphi_1 = \varphi_2 = 46:25°$ with the incident and outgoing light. These angles are used as input parameters in the optical system model. Hence, we obtain the total reflected Mueller matrix of a composite optical system, which consists of cascaded optical elements stacked between the source (polarization state generator) and detector (polarization state analyzer) with Stokes vectors given by the formula:

$$\vec{S}_{out} = M_{Measured}^R \vec{S}_{in} \qquad \text{S4}$$

In our experimental set-up, we employed a *J. A. Woollam* spectroscopic ellipsometer with two rotating compensators (RC2 ellipsometer model) that enables to collect all 16 Mueller matrix elements.

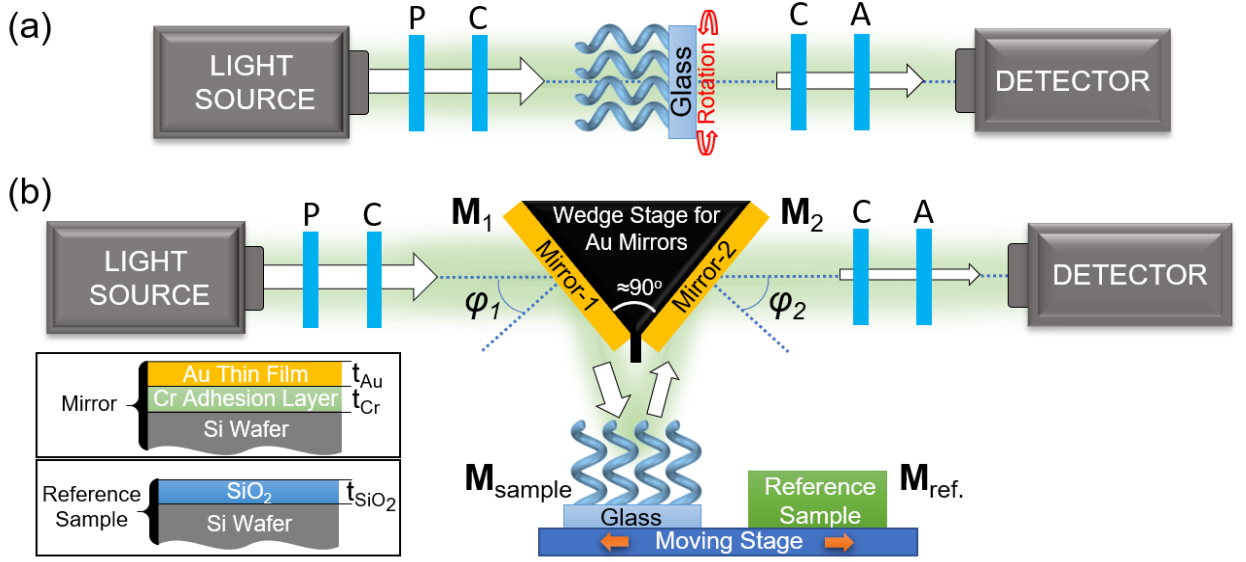

**Figure S3** Schematics of the spectroscopic ellipsometry-based optical set-ups for extracting the Mueller matrix elements in both (a) transmission and (b) reflection mode for normal angle of incidence. The inset schematics in left lower side of (b) show the different thin film layers grown on the Si wafer substrate to implement the mirror and reference sample designs. The optical elements of polarizer (P), compensator (C), and analyzer (A) are also depicted in these experimental set-ups.

The RC2 ellipsometer continuously varies $\vec{S}_{in}$ while observing $\vec{S}_{out}$ to determine the matrix $M^R_{Measured}$. Our sample consists of cascaded optical elements, but we are only interested to the effect of the sample on the incident beam of light. We should therefore isolate and correct the effects of the ancillary optical components on the reflected Mueller matrix computed by the RC2 ellipsometer. Hence, the total measured reflected Mueller matrix is

$$M^R_{Measured} = O_{post} M^R_{Sample} O_{pre}, \qquad \text{S5}$$

where $O_{post}$ and $O_{pre}$ are the optical system Mueller matrices before and after the sample given by:

$$\begin{aligned} O_{pre} &= R_2 M_{Mirror_1} R_1 \\ O_{post} &= R_4 M_{Mirror_2} R_3 \end{aligned} \qquad \text{S6}$$

The 4x4 coordinate rotation matrix $\mathbf{R}_n$ (n=1,2,3,4) is equal to:

$$R_n = \begin{bmatrix} 1 & 0 & 0 & 0 \\ 0 & \cos(2\theta_n) & \sin(2\theta_n) & 0 \\ 0 & -\sin(2\theta_n) & \cos(2\theta_n) & 0 \\ 0 & 0 & 0 & 1 \end{bmatrix}, \qquad \text{S7}$$

where $R_1$ and $R_4$ account for the rotation of the first and second mirror plane relative to the incident plane of the ellipsometer and therefore, $\theta_1 = -\theta_4 \approx 90°$. The minor rotation of the incident plane between the sample and each mirror is accounted by $R_2$ and $R_3$ and it is assumed $\theta_2 = -\theta_3 \approx 0°$. The reflection of each gold mirror at the chosen incident angle is computed by $M_{Mirror_1}$ and $M_{Mirror_2}$. These are derived by the optical constants computed by the reflectance measurements at multiple angles of incidence from each mirror. The unknown $M_{Sample}^R$ matrix is then determined by the measured $M_{Measured}^R$ matrix by inverting the matrices $O_{post}$ and $O_{pre}$ and carrying out the

following matrix multiplication:

$$M_{Sample}^R = O_{post}^{-1} M_{Measured}^R O_{pre}^{-1}, \qquad \text{S8}$$

Note that all measurements are performed with respect to a silicon (Si) reference wafer and the attenuation of the measured Mueller matrices is always normalized to the spectral absorbance of this Si wafer. To compensate for this effect, all Mueller matrix elements are then denormalized by multiplying with a scalar quantity at each wavelength. The denormalization scalar is a (wavelength-by-wavelength) fit parameter that is computed during the calibration process. To fit this denormalization scalar, a reference sample with a different spectral attenuation than the Si wafer is also used, made of a 500 nm thick $SiO_2$ layer grown on the Si wafer. Gold thin film layers ($t_{Au} \approx 100 \ nm$) and ultra thin Chromium (Cr) adhesion layers ($t_{Cr} \approx 10 \ nm$) are deposited on low-doped (100) oriented silicon wafers by using a DC magnetron sputtering system (ATC-2000F sputtering system purchased from AJA International) to create the Au mirrors shown in Figure S3b.

The SiO$_2$ thin film ($t_{\text{SiO}_2} \approx 489\ nm$) reference sample was fabricated on a low-doped (100) oriented silicon wafer by using the RF sputtering method. The corresponding dielectric functions of the Au mirrors and SiO$_2$ thin film reference sample are obtained by the spectroscopic ellipsometry data acquired within the spectral range from 0.72 eV to 6.4 eV at different angles of incidence ranging from 45º to 75º with a 5º step by using the dual rotating compensator ellipsometer (RC2 Woollam model). The extracted optical constants are also employed in the optical system model analysis. To the best of our knowledge, an experimental set-up to directly measure the reflection at normal incidence in terms of Mueller matrix elements has not been presented before in the literature.

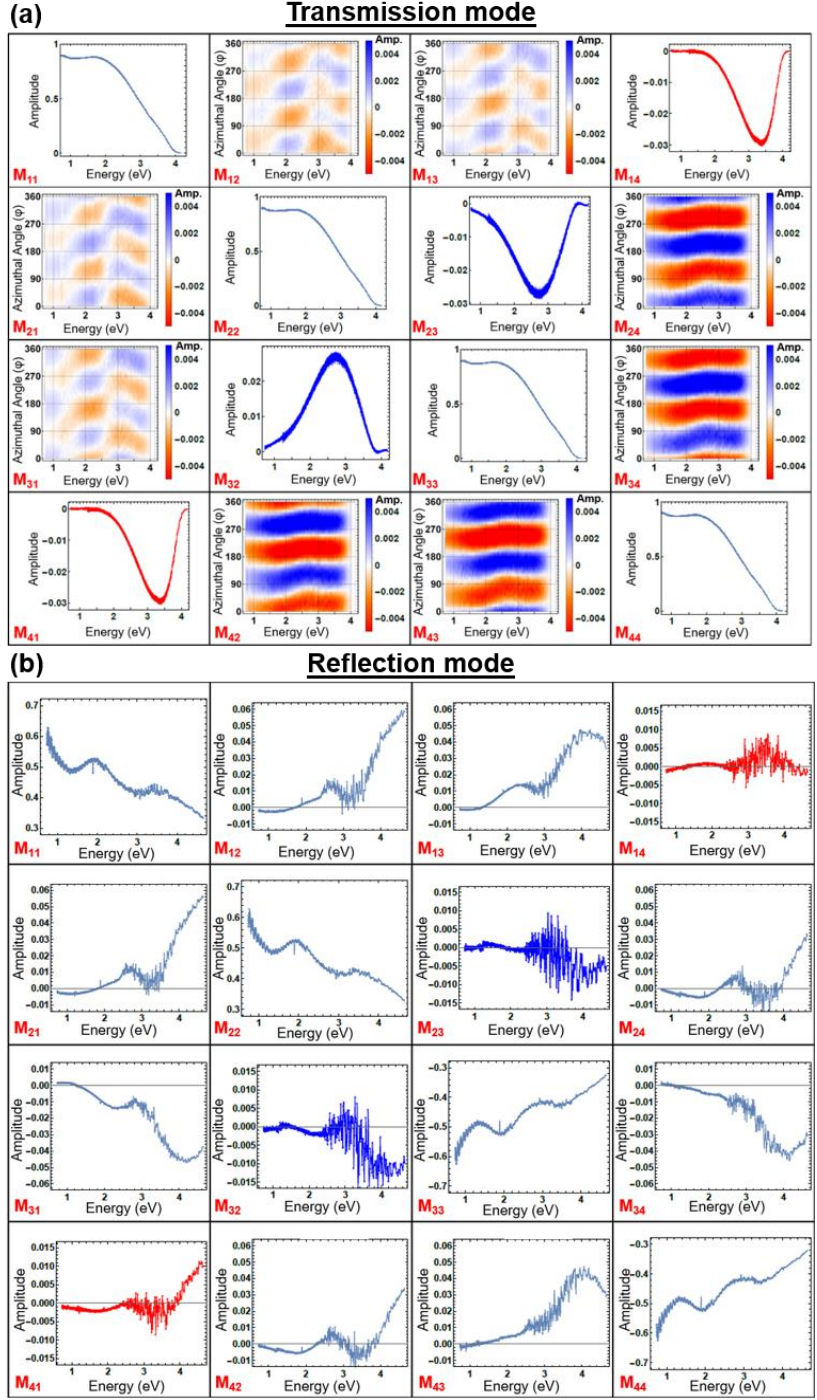

**Figure S4.** (a) Mueller matrix elements spectra of all-dielectric Si helical metamaterial with two turns measured in transmission mode at normal incidence. The major and minor diagonal transmission Mueller matrix elements ($M_{11}$, $M_{22}$, $M_{33}$, $M_{44}$, $M_{41}$, $M_{32}$, $M_{23}$, and $M_{14}$) do not depend on the sample's azimuthal angle. The rest of the transmission elements ($M_{21}$, $M_{31}$, $M_{12}$, $M_{13}$, $M_{42}$, $M_{43}$, $M_{24}$, and $M_{34}$) are shown in (a) as color density plots to demonstrate their change as a function of the azimuthal rotation (from 0° to 360°). (b) The same Mueller matrix elements spectra measured in reflection mode at an arbitrary azimuthal orientation by using the experimental set-up shown in Figure S3b.

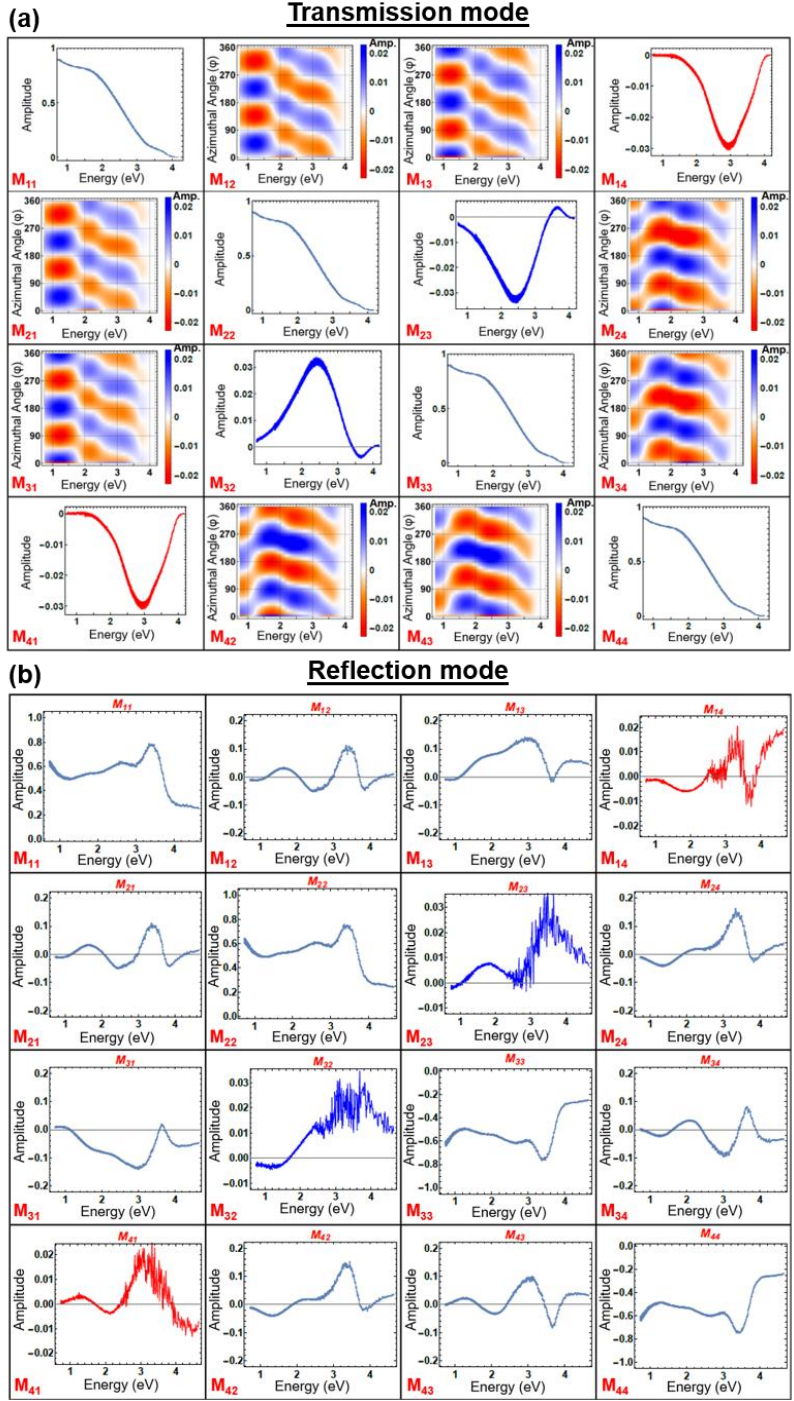

**Figure S5.** (a) Mueller matrix elements spectra of all-dielectric Si-Ag helical metamaterial with two turns measured in transmission mode at normal incidence. The major and minor diagonal transmission Mueller matrix elements ($M_{11}$, $M_{22}$, $M_{33}$, $M_{44}$, $M_{41}$, $M_{32}$, $M_{23}$, and $M_{14}$) do not depend on the sample's azimuthal angle. The rest of the transmission elements ($M_{21}$, $M_{31}$, $M_{12}$, $M_{13}$, $M_{42}$, $M_{43}$, $M_{24}$, and $M_{34}$) are shown in (a) as color density plots to demonstrate their change as a function of the azimuthal rotation (from 0° to 360°). (b) The same Mueller matrix elements spectra measured in reflection mode at an arbitrary azimuthal orientation by using the experimental set-up shown in Figure S3b.

## 3. Additional simulations

We also present the results from additional FEM-based simulations to further elucidate the enhanced chiral dynamics of our new dielectric and plasmonic helical metamaterials with results shown in Figure S6, S7, S8, S9 and S10. More specifically, the change in the theoretically computed Kuhn's dissymmetry factor spectra as a function of various geometrical parameters of the two-turn hybrid Si-Ag helical metamaterial is demonstrated in Figure S6.

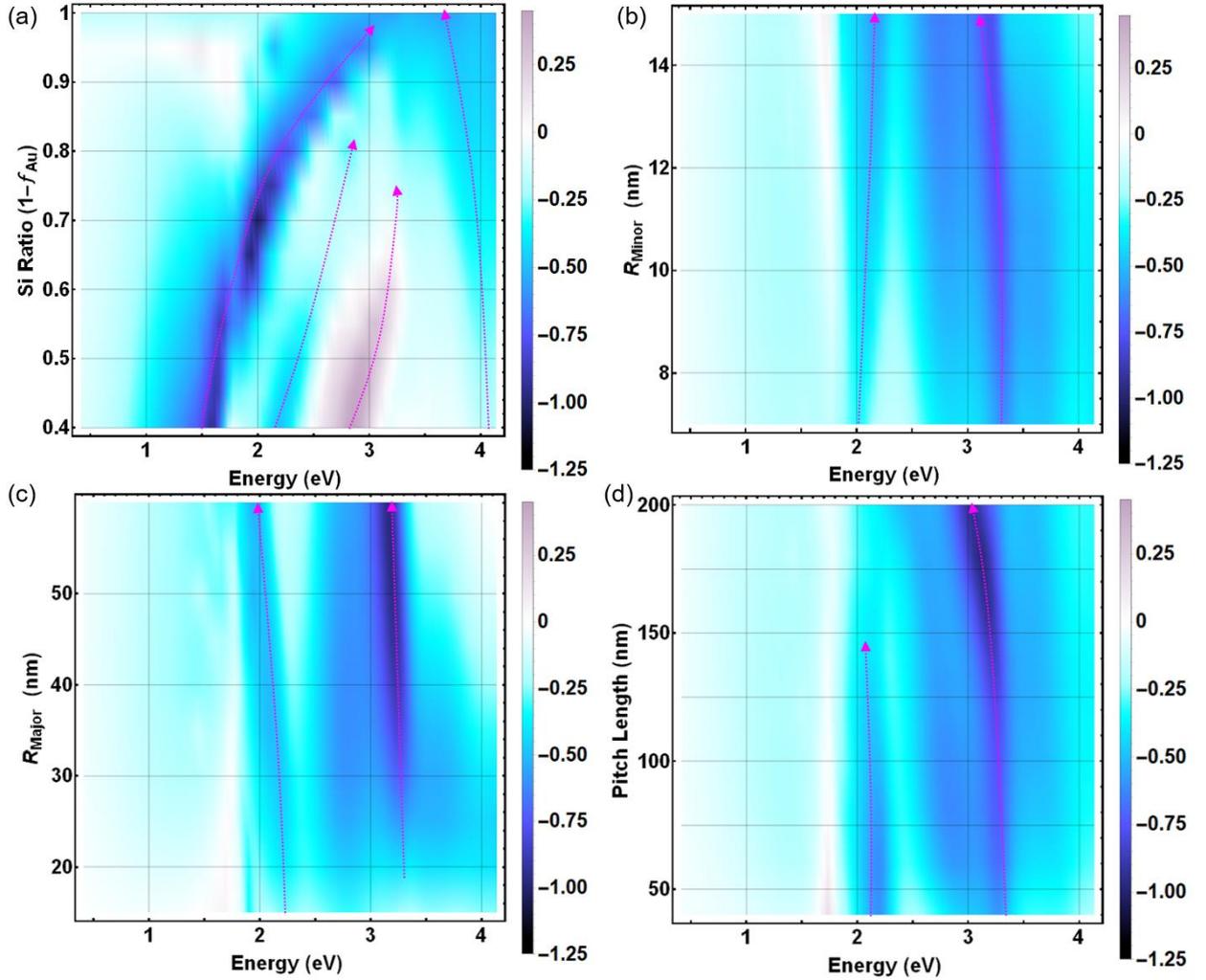

**Figure S6.** Theoretically computed Kuhn's dissymmetry factors, $g_K$, based on FEM simulations as a function of different geometrical parameters: (a) Si ratio ($f_{Si} = 1 - f_{Ag}$), (b) minor radius ($R_{minor}$), (c) major radius ($R_{major}$), and (d) pitch length ($A_{pitch}$) plotted for two turn Si-Ag helical metamaterials.

In addition, Figure S7 shows the Kuhn's dissymmetry factor spectra of different helical metamaterials when the Si material is assumed to be single crystalline, different than the polycrystalline Si properties used in the simulations of the main paper. Figure S7a demonstrates the spectral evolution of $g_K$ as a function of number of turns for the single crystalline Si helical metamaterial. The emergence of a single dip in the g-factor spectra is observed and its maximum amplitude is found to be $g_K$=1.05 at λ= 393nm when the turn number of is chosen to be four. When the dielectric permitivity constant of polycrystalline Si is used in the simulations in the main paper, the dip in the $g_K$ spectra is lower and its amplitude is equal to 0.4 at λ = 548nm for the Si helical metamaterial with four turns. Figure S7b shows the effect of number of turns on the evolution of the $g_K$ spectra in the case of Si-Ag helical metamaterials. In this case, we observe a narrow g-factor with a dip at 540 nm, where its amplitude is equal to 1.05, and there is almost no spectral shift as a function of the turn number. To achieve spectral tunability in the $g_K$ spectra resonances, we also performed the study of increasing the number of Ag subsegments in each helix and used a fixed number of two turns. With this study, we inspect how the increase in the number of metal-dielectric interfaces results in an enhanced localized surface plasmon resonance effect and therefore achieves spectral tunability in the $g_K$ dip. As it is seen from Figure S7c, the increased total Si ratio (i.e., the decrease in number of subsegments) results in a significant blue spectral shift in the $g_K$ dip from 2.3eV to 3.25eV. The single crystalline Si dielectric properties are used only in Figure S7 and all the other theoretically obtained Figure utilize the polycrystalline Si properties.

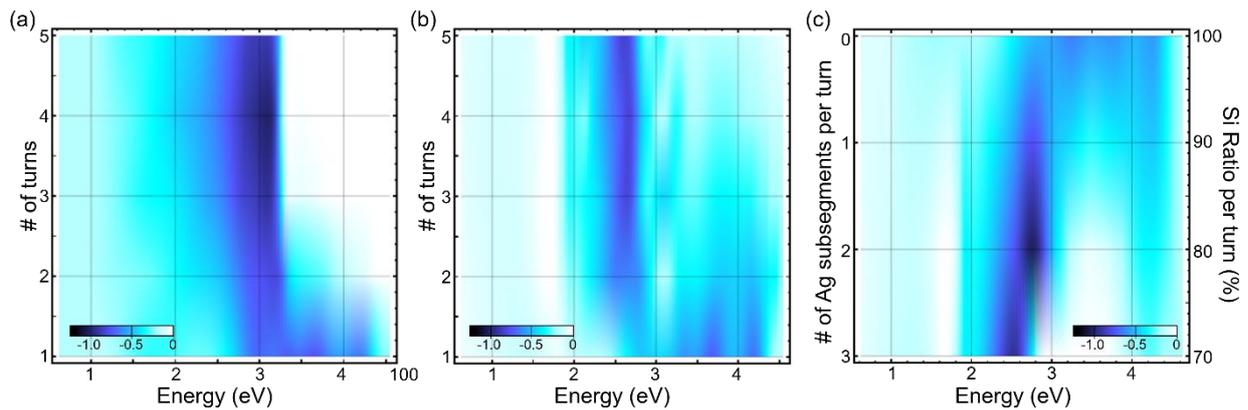

**Figure S7** Theoretically computed Kuhn's dissymmetry factors, $g_K$, based on FEM simulations as a function of number of turns for (a) Si and (b) Si-Ag helical metamaterials, and (c) number of Ag subsegments for two turn plasmonic Si-Ag helical metamaterials. All these theoretical results were obtained by using the single crystalline Si dielectric permittivity properties.

The localized surface plasmon resonance effects are depicted more clear in Figure S8a and b, where it is demonstrated that the incorporation of multiple Ag helical subsegments leads to tunable g-factor response due to the multiple metal-dielectric interfaces with enhanced electric fields. More specifically, an evanescently decaying electric field enhancement is observed along each interface, as clearly shown in Figure S8b plotted for the points (1)-(3) in the g-factor distribution depicted in Figure S8a. This enhanced field due to plasmonic effects can serve as an excellent condition to improve chiral sensing applications. The complete evolution of electric and magnetic field distributions as a function of number of Ag subsegments per turn for various two turn Si-Ag helical metamaterials is shown in Figure S9. The resonant spectral locations of each plot in Figure S9 are chosen from Figure S8a. The localized surface plasmon resonance causes the incident beam to be spatially confined to those extremely subwavelength interfaces, leading to strong field enhancement. This enhancement of the incident electric field is the main source of tunability in the dissymmetry factor of the Si-Ag helical metamaterials. As a result of the strong enantioselective property of the proposed hybrid metamaterials that can be tuned by increasing the plasmonic helical subsection area, a macroscopically distinguishable change in the color (or frequency) of the transmitted circular polarized light is expected to occur. The resulting colors of transmitted light shown in Figure S8c are extracted from the standardized tristimulus system. This system is also known as the CIE (International Commission on Illumination) color system, which establishes a framework for visually matching a color under standardized conditions against the three primary colors (red, green, and blue). The resulting primary three colors are expressed as positive X, Y, and Z values, also known as tristimulus values.[7] These colors are computed by simulations and are arranged into a matrix formation shown in Figure 8c, where the columns demonstrate a set of colors as a three-slice pie-chart. The top left and right colors are the transmitted light colors when the circular polarization state of the incident light is chosen to be either RH or LH, respectively. The bottom color is derived by the difference between the transmitted RH and LH circular polarized light. This theoretical study clearly demonstrates how the spectral tunability of the chiroptical response evolves depending on both the number of turns (columns) and metallic subsegments per turn (rows). Hence, strong and tunable chiral response is derived with increased tunability as the number of helical turns or plasmonic subsegments increases.

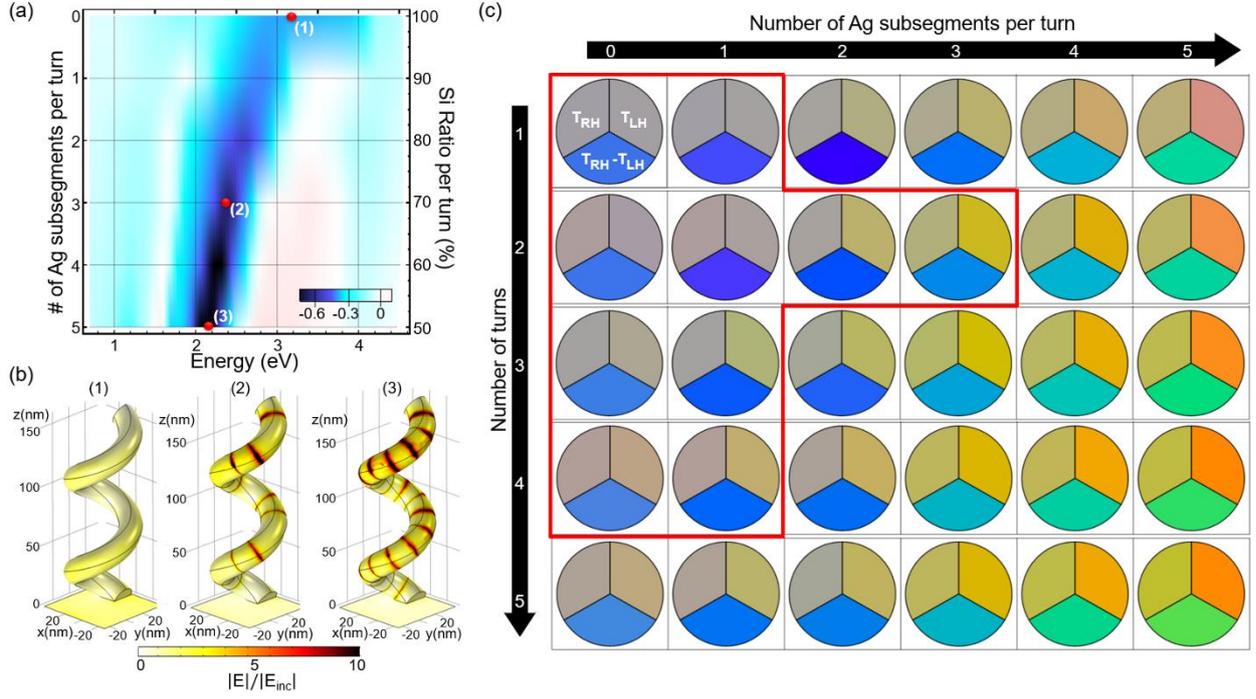

**Figure S8**. (a) The spectral evolution of Kuhn's dissymmetry factor as a function of the number of Ag subsegments per turn in the case of two turn hybrid Si-Ag plasmonic helical metamaterials. (b) The incident light is chosen to be RH circular polarized and the calculated average electric field magnitude ($|\vec{E}| = \sqrt{E_x^2 + E_y^2 + E_z^2}$) on the surface of the plasmonic Si-Ag nanohelix at points (1)-(3) in caption (a) is normalized to the magnitude of the incident electric field to compute the field enhancement. (c) Matrix of three slice pie-chart forming an illustrative infographic to demonstrate the tunability of the presented hybrid plasmonic helical metamaterials. Each color represents light transmitted through the various helical metamaterials when the polarization state of the incident light is either LH ($T_{LH}$) or RH ($T_{RH}$). The color of the transmission spectra difference ($T_{RH}$ -$T_{LH}$) is also depicted. The evolution of colors is presented as a function of the Ag subsegments number per turn (columns) and number of turns (rows). The red line indicates metamaterials whose g-factor values are shown in Figure 2-4 in the main paper.

Finally, Figure S10a and b show the Kuhn's dissymmetry factor spectrum of (a) LH and (b) RH Si-Ag helical metamaterials as a function of the Si subsegment ratio ranging from 40% to 100%. As expected, the computed $g_K$ color density distributions of the different handedness helical metamaterials are mirror symmetric to each other. The change in Si ratio leads to multiple resonances in the $g_K$ spectrum and the real part of the electric field distributions at those resonances are plotted in Figure S10c.

In order to further investigate the chiroptical properties of helical heterostructures, we performed another systematic study to unravel the spectral tunability of chirality as a function of turn number and Si ratio of each helical Si subsegment. Figure S10d shows the computed colors of RH ($T_{RH}$) and LH ($T_{LH}$) transmitted light and also the difference between them ($T_{RH}$-$T_{LH}$) for RH Si and Si-Ag helical metamaterials in a pie chart matrix. While the columns of the pie-chart matrix show the color evolution when the Si rate is decreased from 1 to 0.4, the rows of the pie-chart matrix demonstrate the color evolution as a function of turn number (from one to five). Again, it is proven that increased plasmonic Ag subsegments will lead to a more tunable and enhanced chiral response.

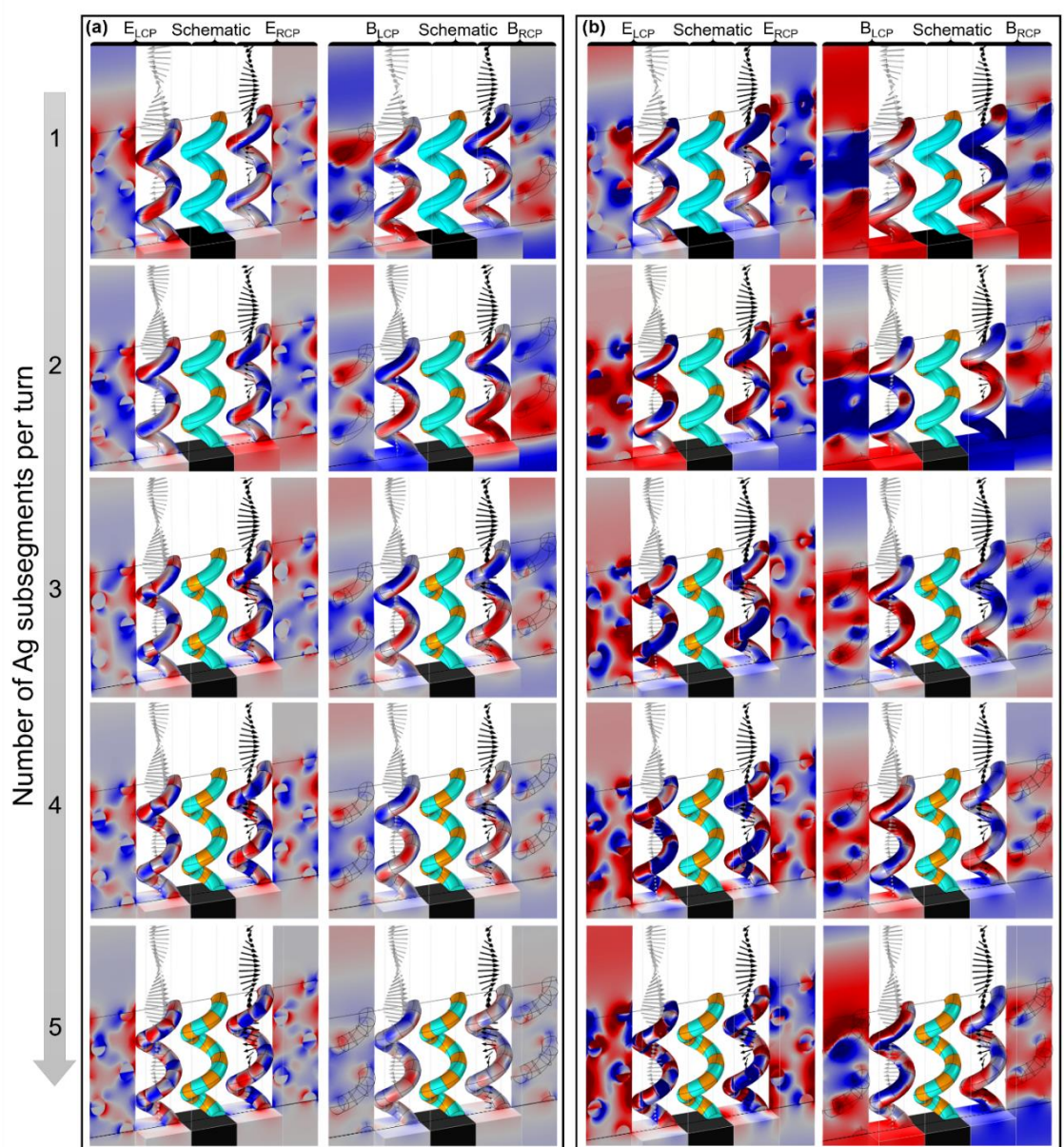

**Figure S9.** Evolution of electric and magnetic field distributions as a function of number of Ag subsegments per turn for various two turn Si-Ag helical metamaterials. The resonant spectral locations were chosen from Figure S8a. (a) The first mode spectral location evolves from ≈2.17 eV (10 subsegments /turn) to ≈3.08 eV (2 subsegment/turn) (b) The second mode with much lower g-factor evolves from 4.2 eV (10 subsegments/turn) to 4.1 eV (2 subsegments). The first panel (first column) at each row from left to right represents a 2D cut-slice color density plot of electric field (when the incident light polarization is RH), surface color density plot of electric field (when the incident light polarization is RH), schematic illustration of helical metamaterial, 3D surface color density plot of the electric field (when the incident light circular polarization is LH), and 2D cut-slice color density plot of electric field (when the incident light circular polarization is LH). The second panel (second column) at each row from left to right demonstrates a 2D cut-slice color density plot of magnetic field (when the incident light circular polarization is RH), surface color density plot of magnetic field (when the incident light circular polarization is RH), schematic illustration of helical metamaterial, 3D surface color density plot of the magnetic field (when the incident light circular polarization is LH), and 2D cut-slice color density plot of magnetic field (when the incident light circular polarization is LH).

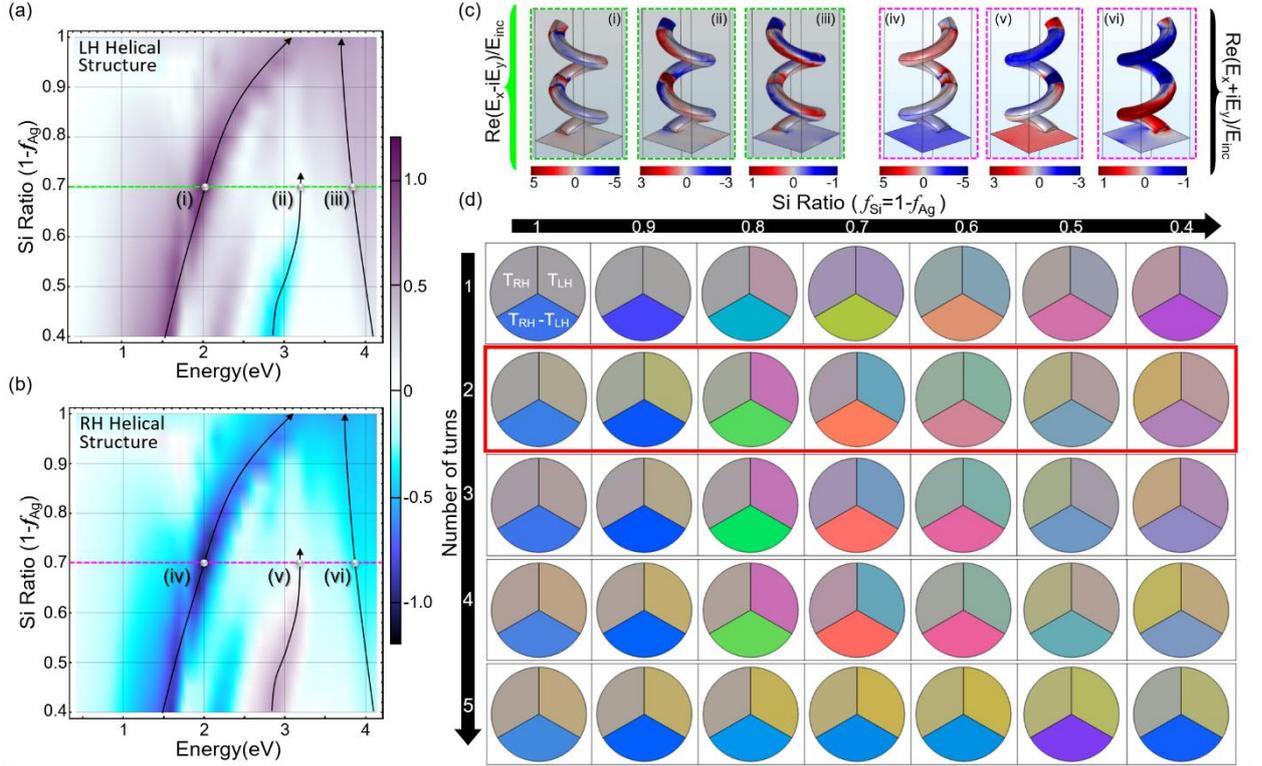

**Figure S10.** Theoretically computed Kuhn's dissymmetry factors, $g_K$, based on FEM simulations as a function of the Si ratio for (a) LH and (b) RH two turn Si-Ag helical metamaterials. (c) Surface color density plots of the real part of E and B fields at the modes depicted as (i), (ii), (iii), (iv), (v), and (vi) in the g-factor spectra in captions (a)-(b). (d) Pie chart matrix to demonstrate the resulting color of transmitted light in RH (top left section of each pie chart), LH (top right section of each pie chart), and the difference between them (bottom section of each pie chart) for Si and Si-Ag helical metamaterials. The matrix columns show the color evolution when the Si ratio is decreased from 1 to 0.4 and the matrix rows depict the color evolution as a function of turn number (from one to five). The red box part of the pie chart matrix corresponds to the two turn RH Si-Ag helical metamaterial with g-factor presented in caption (b).

## 4. Optical helicity

The ability of a nanostructure to interact or, equivalently, sense chiral molecules can be characterized by computing the optical helicity (OH), also known as optical chirality, which is given by the formula: $OH = \left(\frac{\mu_o \varepsilon_o}{2}\right)\left(\vec{E} \cdot \vec{H}^*\right)$, where $\mu_o$ and $\varepsilon_o$ are the magnetic permeability and electric permittivity of free space and $\vec{E}$ and $\vec{H}$ are the induced electric and magnetic fields when circular polarization is applied.[8]-[9] We compute the optical helicity for RH ($OH_{RH}$) and LH ($OH_{LH}$) circular polarized incident radiation by using FEM-based simulations in the same frequency points where ΔE and ΔB were computed in the main paper (see Figure 2c and Figure 3c). The net change ($\varDelta OH$) between $OH_{RH}$ and $OH_{LH}$ can characterize the ability of our metamaterials to sense chiral molecules and is shown in Figure S11. More specifically, the color density distributions of the net change of the optical helicity ($\varDelta OH$) for both all-dielectric Si and Si-Ag helical metamaterials with four turns are shown in Figures S11a and S11b, respectively. Interestingly, in the vicinity of the Ag subsegments (Figure S11b), we observe strong negative $\varDelta OH$ values (red color) for the hybrid plasmonic Si-Ag metamaterial, while in the case of the all-dielectric Si helical metamaterial (Figure S11a), we only observe positive values (light blue color), which are lower compared to the plasmonic metamaterial. These results directly prove that the presented hybrid plasmonic metamaterials can be used as efficient chiral molecular sensors.

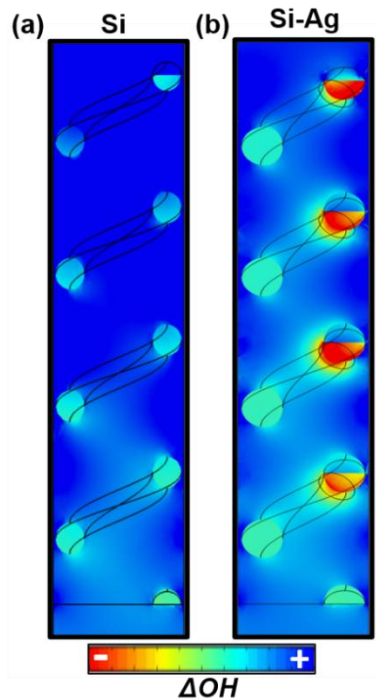

**Figure S11.** The net change ($\varDelta OH$) between $OH_{RH}$ and $OH_{LH}$ plotted on a 2D slice for (a) all-dielectric Si and (b) hybrid plasmonic Si-Ag helical metamaterials with four turns monitored at the same photon energy with Figure 2c and Figure 3c in the main paper.

# 5. Transmission and reflection dissymmetry factors

The transmission-based dissymmetry factor ($g_{Trans}$) is computed by using the definition provided in [10], which was also used in other relevant works.[11]-[12] We compute $g_{Trans}$ in terms of the measured Mueller matrix elements by using the formulas given by Eqs. S2:

$$g_{Trans} = 2\left(\frac{T_+ - T_-}{T_+ + T_-}\right) = 2\left(\frac{M_{14}^{Trans}}{M_{11}^{Trans}}\right) \qquad \text{S9}$$

The reflection-based dissymmetry factor ($g_{Ref}$) is computed by following a similar approach but now using the Mueller matrix elements measured in reflection:

$$g_{Ref} = 2\left(\frac{R_+ - R_-}{R_+ + R_-}\right) = 2\left(\frac{M_{14}^{Ref}}{M_{11}^{Ref}}\right) \qquad \text{S10}$$

The computed reflection and transmission-based g-factor spectra are shown in Figure S12, where the results are obtained for all-dielectric Si (S12a and S12b) and plasmonic hybrid Si-Ag (S12c and S12d) helical metamaterials with varying turns (one to four). It is interesting to note that the high transmission-based g-factor values obtained at 3.5eV in Figure S12a, corresponding to the all-dielectric Si helical metamaterial with four turns, approach the ultimate limit of plus or minus two in g-factor values. This is by far the highest ever reported transmission-based g-factor value in the literature, at least to our knowledge, compared to any relevant chiral nanostructure operating at the same frequency range. Moreover, the increased ohmic losses due to the plasmonic subsegments decrease the transmission-based g-factor amplitude, as can be clearly seen in Figure S12c. However, interestingly, in this plasmonic structure an additional dip in the g-factor spectra is obtained (Figure S12c), which can be made tunable by increasing the number of helical turns. This additional dip is attributed to the excitation of localized surface plasmon resonances along the metallic subsegments of the hybrid helix.

Note that Figures S12b and S12d clearly demonstrate that the chiral response of the currently presented metamaterials is much weaker in reflection compared to transmission. Hence, the use of the currently presented GLAD nanostructures as efficient chiral optical filters is only advisable for transmission, since the chiral effect is much weaker in reflection. In addition, it is worthwhile to mention that, for the case of hybrid plasmonic Si-Ag metamaterials (Figure S12d), the reflection-based g-factor is higher compared to the all-dielectric nanostructures (Figure S12b) because more light is reflected by the plasmonic structures due to the increased mismatch with the surrounding environment introduced by the metal content. These conclusions are inline to the reflection measurements shown in the next section 6. Finally, it is important to stress that the noise in the reflectance data (Figures S12b and S12d) is not an inherent property of our GLAD fabricated metamaterials, but is due to insufficient probing light intensity in our reflection measurement system mainly because the reflection signals from our metamaterials are very weak, a problem that is even more acute in higher frequencies.

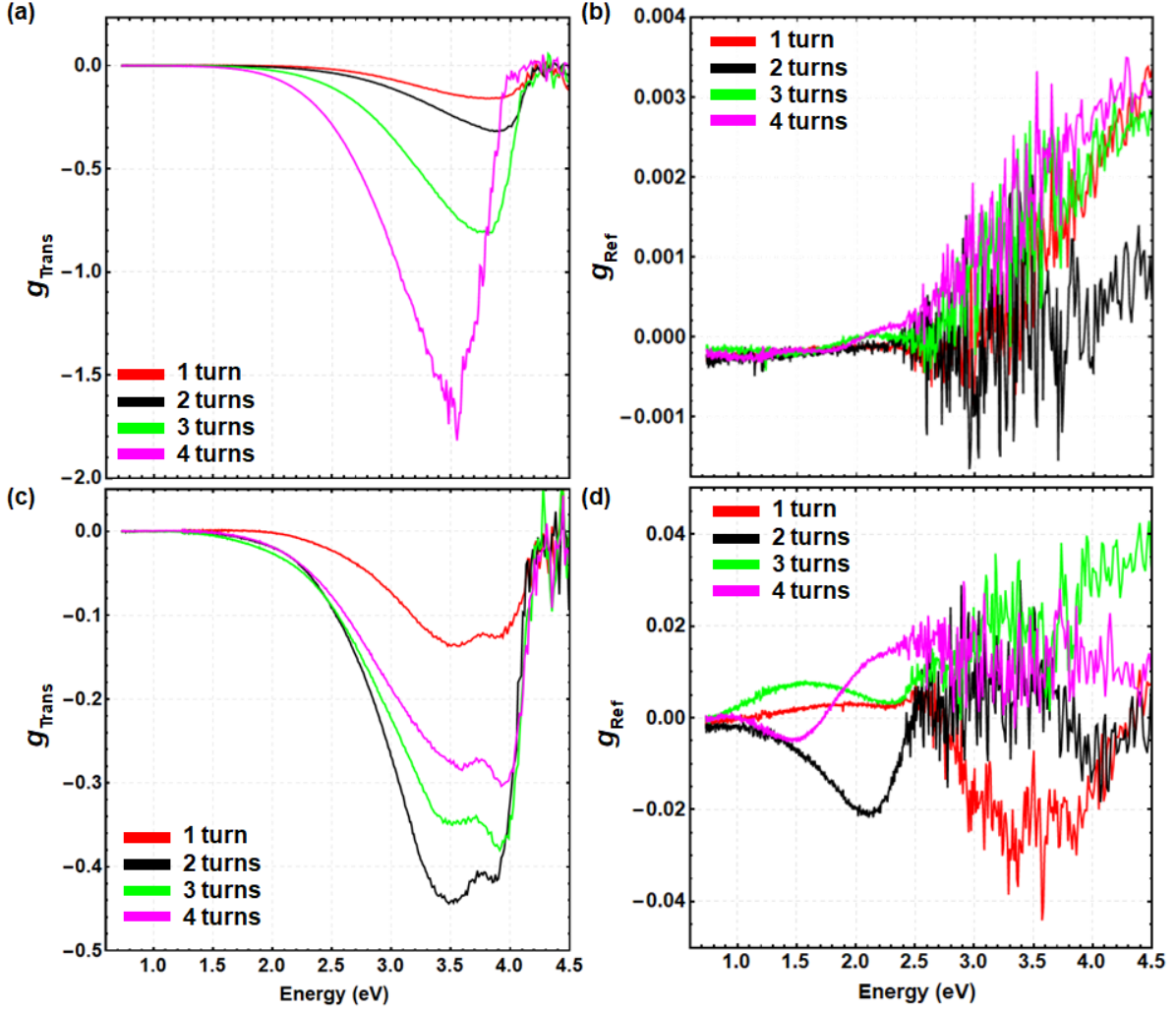

**Figure S12.** Experimentally measured spectral evolution of (a), (c) transmission and (b), (d) reflection-based dissymmetry factors for (a), (b) all-dielectric Si and (c), (d) plasmonic hybrid Si-Ag helical metamaterials with varying turns (one to four).

## 6. Circular polarized transmission and reflection measurements

Next, we use the relations shown in Eqs. S2 and obtain the spectral evolution of the reflectance ($R_\pm$), transmittance ($T_\pm$), and absorbance ($A_\pm = 1 - R_\pm - T_\pm$) for both RH (+) and LH (-) circular polarized incident radiation by using the Mueller matrix elements measured in transmission and reflection (Figures S4 and S5). Figure S13 shows the measured $R_\pm$, $T_\pm$, and $A_\pm$ spectra for all-dielectric Si (Figures S13a-S13c) and hybrid plasmonic Si-Ag (Figures S13d-S13f) helical metamaterials with four turns. We choose to present only the four turn designs, since these metamaterials have the more pronounced chiral response. Interestingly, the transmittance ($T_\pm$) and absorbance ($A_\pm$) are different for LH and RH polarized radiation in a very broad frequency range, inline to the broadband g-factor results presented in the main paper. Hence, the broadband chiral response of our metamaterials is proven by these plots. The broadband chiral response is a unique feature of the currently presented metamaterials, attributed to individual (from each unit cell) resonances combined with collective responses or photonic bands (due to mutual coupling from neighboring periodic unit cells) in the case of both metallic[13]-[16] and dielectric[17]-[20] 3D helical metamaterials. Moreover, the reflection response shown in Figures S13b and S13e is much weaker compared to the transmission results, which confirms the g-factor calculations presented in the previous section. Hence, the currently presented GLAD metamaterials can be used as efficient chiral optical filters only in transmission, since the chiral effect is much weaker in reflection. In addition, the weak reflection is further reduced in higher frequencies and this causes increased noise to our detected reflected signals, a problem that was also mentioned in the previous section. Finally, the weak reflection peaks at smaller frequencies in Figures S13b and S13e are product of constructive interference due to the finite thickness of our samples that can lead to mismatch with the surrounding environment which can be further enhanced by localized surface plasmon resonances, as discussed in the previous section.

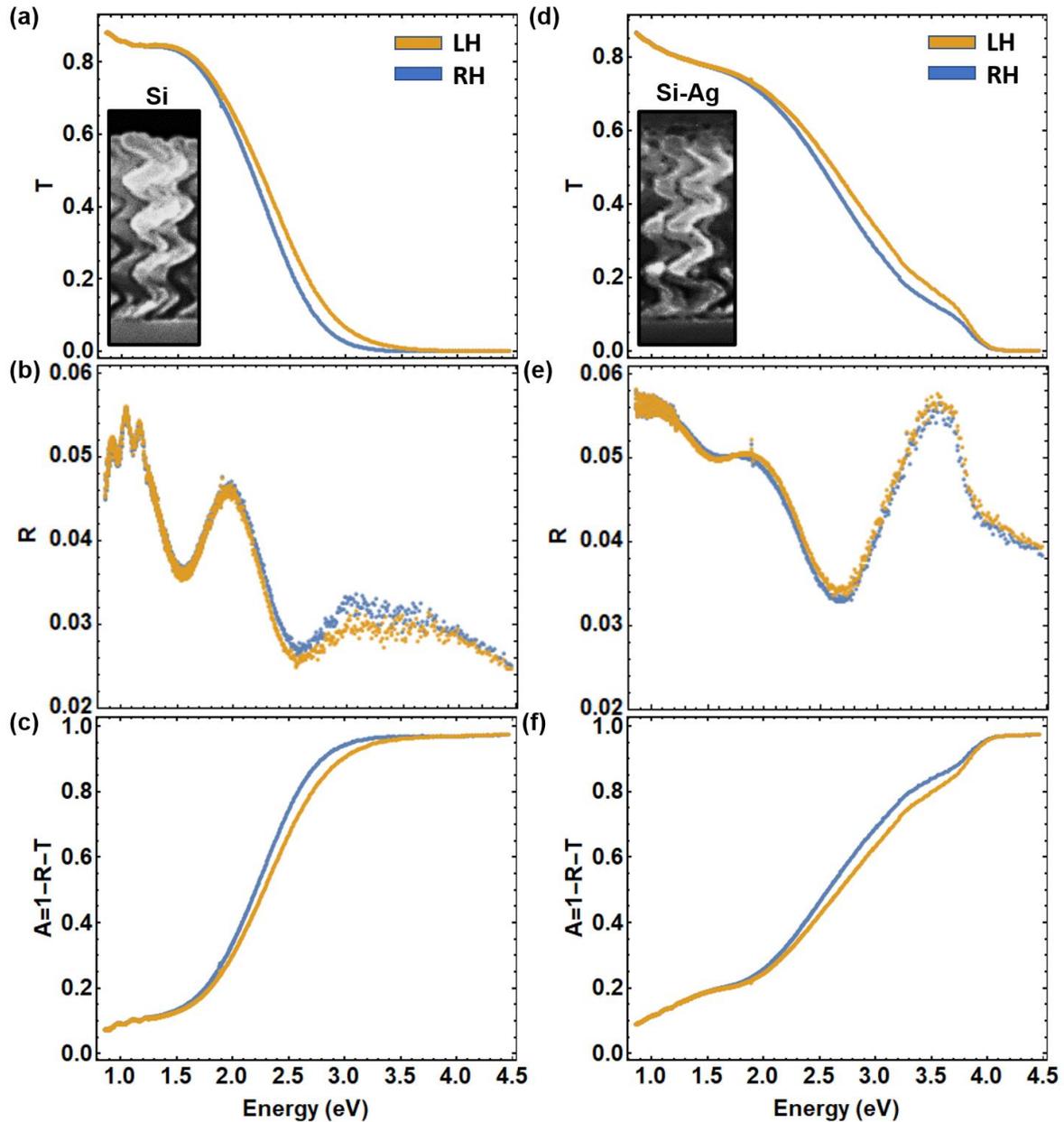

**Figure S13.** Experimentally measured transmittance, reflectance, and absorbance responses for LH and RH circular polarized light exciting (a)-(c) all-dielectric Si and (d)-(f) hybrid plasmonic Si-Ag helical metamaterials with four turns. SEM images of single Si and Si-Ag helical nanostructures are inlaid to their corresponding transmittance spectra.